\documentclass[acmtog, nonacm=true]{acmart}
\acmSubmissionID{0130}

\usepackage{booktabs} 

\usepackage[ruled]{algorithm2e} 

\SetAlFnt{\small}
\SetAlCapFnt{\small}
\SetAlCapNameFnt{\small}
\SetAlCapHSkip{0pt}

\usepackage{graphicx}
\usepackage{grffile}
\usepackage{capt-of}
\usepackage{xcolor}
\usepackage{amsmath,amsfonts,dsfont,pifont,bm,bbm,mathrsfs,mathtools,nicefrac} 
\usepackage{algpseudocode,listings} 
\usepackage{booktabs,multirow,adjustbox,diagbox,threeparttable}
\usepackage{subcaption}
\usepackage{multirow}
\usepackage{array}
\newcolumntype{L}[1]{>{\raggedright\let\newline\\\arraybackslash\hspace{0pt}}m{#1}}
\newcolumntype{C}[1]{>{\centering\let\newline\\\arraybackslash\hspace{0pt}}m{#1}}
\newcolumntype{R}[1]{>{\raggedleft\let\newline\\\arraybackslash\hspace{0pt}}m{#1}}

\usepackage{enumitem}
\setlist[enumerate]{itemsep=2pt, topsep=10pt}

\newcommand{\revised}[1]{#1}

\acmJournal{TOG}

\usepackage{fp}
\usepackage{expl3}[2012-07-08]
\ExplSyntaxOn
\ExplSyntaxOff

\usepackage{amsmath,amsfonts,bm}

\newcommand{\ours}{GroomGen}

\newcommand{\ignorethis}[1]{}

\def\eqref#1{equation~\ref{#1}}

\def\1{\bm{1}}

\def\vb{{\bm{b}}}

\def\vd{{\bm{d}}}

\def\vh{{\bm{h}}}

\def\vl{{\bm{l}}}
\def\vm{{\bm{m}}}
\def\vn{{\bm{n}}}

\def\vp{{\bm{p}}}

\DeclareMathAlphabet{\mathsfit}{\encodingdefault}{\sfdefault}{m}{sl}
\SetMathAlphabet{\mathsfit}{bold}{\encodingdefault}{\sfdefault}{bx}{n}

\newcommand{\ignore}[1]{}

\begin{document}

\title{\ours: A High-Quality Generative Hair Model Using Hierarchical Latent Representations}

\author{Yuxiao Zhou}
\affiliation{%
  \institution{ETH Zurich}
  \country{Switzerland}}
\email{yuxiao.zhou@inf.ethz.ch}

\author{Menglei Chai}
\affiliation{%
  \institution{Google Inc.}
  \country{United States of America}}
\email{mengleichai@google.com}

\author{Alessandro Pepe}
\affiliation{%
  \institution{Google Inc.}
  \country{United States of America}}
\email{apepe@google.com}

\author{Markus Gross}
\affiliation{%
  \institution{ETH Zurich}
  \country{Switzerland}}
\email{grossm@inf.ethz.ch}

\author{Thabo Beeler}
\affiliation{%
  \institution{Google Inc.}
  \country{Switzerland}}
\email{tbeeler@google.com}

\begin{abstract}

Despite recent successes in hair acquisition that fits a high-dimensional hair model to a specific input subject, generative hair models, which establish general embedding spaces for encoding, editing, and sampling diverse hairstyles, are way less explored.
In this paper, we present \textit{\ours}, the first generative model designed for hair geometry composed of highly-detailed dense strands.
Our approach is motivated by two key ideas. First, we construct \textit{hair latent spaces} covering both individual strands and hairstyles.
The latent spaces are compact, expressive, and well-constrained for high-quality and diverse sampling.
Second, we adopt a \textit{hierarchical hair representation} that parameterizes a complete hair model to three levels: single strands, sparse guide hairs, and complete dense hairs.
This representation is critical to the compactness of latent spaces, the robustness of training, and the efficiency of inference.
Based on this hierarchical latent representation, our proposed pipeline consists of a \textit{strand-VAE} and a \textit{hairstyle-VAE} that encode an individual strand and a set of guide hairs to their respective latent spaces, and a \textit{hybrid densification step} that populates sparse guide hairs to a dense hair model.
\textit{\ours} not only enables novel hairstyle sampling and plausible hairstyle interpolation, but also supports interactive editing of complex hairstyles, or can serve as strong data-driven prior for hairstyle reconstruction from images.
We demonstrate the superiority of our approach with qualitative examples of diverse sampled hairstyles and quantitative evaluation of generation quality regarding every single component and the entire pipeline.

\end{abstract}

\keywords{Strand-level hair modeling, hairstyle generation}

\begin{teaserfigure}
\includegraphics[width=\linewidth]{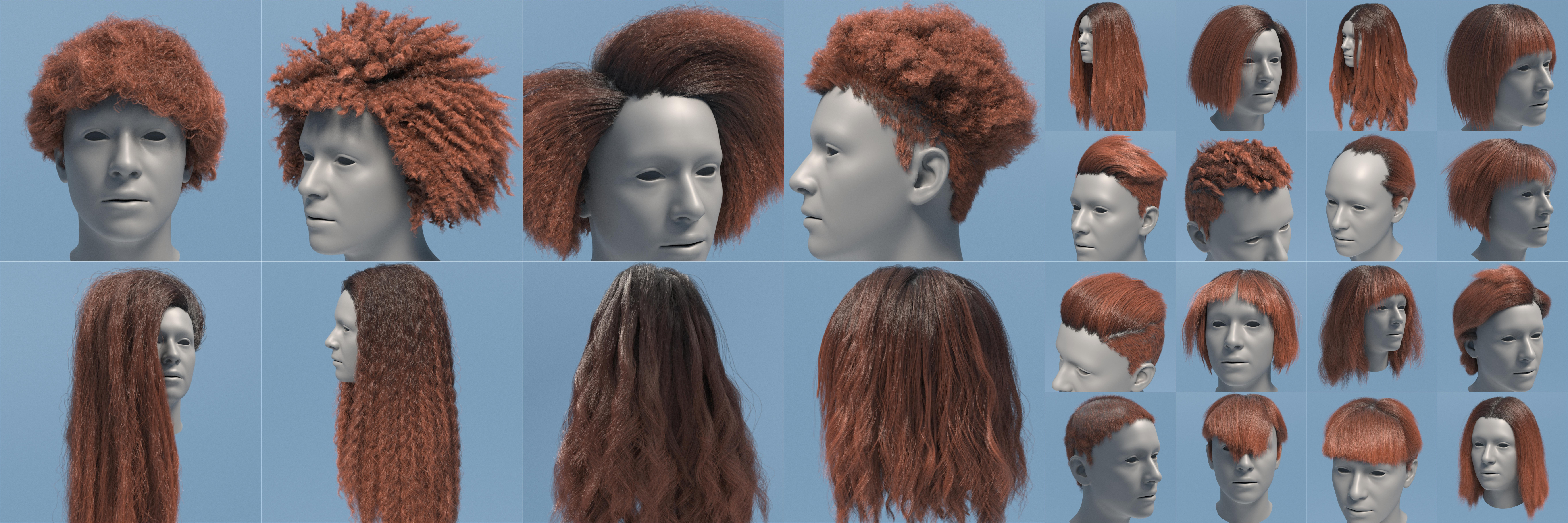}
\caption{Our method is able to automatically generate diverse high-quality hairstyles from random latent vectors.}
\label{fig:teaser}
\end{teaserfigure}

\maketitle

\section{Introduction}
\label{sec:intro}

Hair substantially contributes to a person's appearance, and we frequently change it to express ourselves.
As such it plays a critical role in depicting not just our physical appearance but also reflecting our individuality, mood, and cultural belonging.
Hair digitization and modeling have recently garnered much attention, highlighting the exciting potential of creating high-quality hairstyles that contribute significantly to the perceived realism of virtual human avatars.
However, unlike any other parts of ourselves, such as faces, bodies, or hands, hair geometry is highly intricate and unstructured, making it exceptionally challenging to represent or model.

Empirically, the complexity of hair arises from two main levels.
Locally, each individual strand corresponds to a 1D curve embedded in 3D space, originating from the scalp, defined by intrinsic properties that give rise to diverse curliness or waviness, and modulated by external conditions such as gravity.
Globally, comprising hundreds of thousands of hair strands, the overall hairstyle exhibits a high-level structure that combines coherency among neighboring strands with independent variations on a per-strand basis.
Given the intricate nature of hair, much of the existing research focuses on hair \textit{acquisition}, which often involves overfitting high-dimensional hair representations, typically Euclidean positions of densely sampled vertices, to various types of inputs (multi-view images~\cite{nam2019strand,winberg2022facial,zhang2017data,beeler2012coupled,luo2012multi}, single photos~\cite{hu2015single,chai2016autohair}, or specialized sensors~\cite{herrera2012lighting}).
While high-fidelity reconstruction for specific subjects is achieved, without proper parameterization that embeds the different hair within a shared compact space, their outputs lack the generalization capability necessary to support interpolation, manipulation, or novel hairstyle synthesis.

On the other hand, \textit{generative hair models}, the main focus of this work, remain relatively unexplored.
The pioneering work on hair geometry synthesis~\cite{wang2009example} proposes a 2D hair embedding that can generate new hairstyles from given exemplars through texture synthesis.
However, due to the inherent limitations of explicit strand geometry encoding and homogeneous texture synthesis, this approach can only handle short to medium-length hair with uniform styles.
More recently, Volumetric Hair VAE~\cite{saito20183d} demonstrates the potential of generating novel hairstyles by interpolating existing ones in a latent volume space.
However, the combination of volumetric flow field and post-processing strand tracing often results in over-smoothed geometry with limited strand-level detail and inter-strand variation.
The objective of our work is to develop a novel architecture for strand-level hair generation, capable of synthesizing diverse hairstyles with high-quality dense geometry in a computation- and memory-efficient manner.
To this end, we aim to establish a new hair representation that is highly compact and efficient, capable of capturing common hairstyles through a shared parameterization, and expressive enough to encompass both global structural characteristics and local fine details.

We introduce \textit{\ours}, a generative model for diverse and high-quality hairstyle synthesis.
Our method is rooted in a hierarchical hair representation, inspired by the conventional practice of guide-hair-based authoring in visual effects.
We represent a hairstyle using three levels of abstraction: strand latent codes for \textit{individual strands}, low-resolution latent-maps for \textit{sparse guide hairs}, and high-resolution strand-maps for \textit{dense hairstyles}.
This hierarchy not only achieves significant compression, robust training, and efficient inference without compromising expressiveness, but also establishes a versatile multi-level embedding space for sampling diverse and valid hairstyles.
Based on this hierarchical representation, we design the entire hair generation pipeline as \revised{three} major components:
\begin{enumerate}[leftmargin=25pt]
\item At the single strand level, we employ a \textit{strand variational autoencoder} (strand-VAE) to establish a low-dimensional latent space for encoding diverse strand geometry.
\item Building upon the strand latent space, our \textit{hairstyle variational autoencoder} (hairstyle-VAE) encodes the sparsely-sampled guide hairs into a hairstyle feature vector.
\item To generate dense hair from sparse guide strands, we propose a GAN-based \textit{neural upsampler} that synthesizes high-resolution hair geometry, followed by a heuristic refinement step that allows user control for customizing details.
\end{enumerate}

Our compact model possesses the capability to represent and generate diverse hairstyles with high visual fidelity.
This versatility makes it useful for a wide range of applications, such as simulation, generating training data for downstream models through (un)conditional sampling, serving as a powerful prior for robust image-based hair reconstruction, and facilitating rapid hairstyle creation and exploration for artists.

\section{Related Work}
\label{sec:related}

\paragraph{Strand Representation of Hairs.}
A common approach to represent a hair model is by using a collection of hair strands, where each strand is defined as a polyline consisting of tens or hundreds of vertices.
While this representation is intuitive and expressive, it often becomes heavy and redundant.
To address this issue, previous works such as~\cite{bertails2005predicting, bertails2006super} propose to use the super-helix as a compact approximation of hair strands.
This representation requires only a few parameters per strand, but the resulting reconstruction is typically over-smoothed.
In the recent work by~\cite{rosu2022neural}, neural representations of hair strands are explored.
The authors adopt the modulated sine network structure~\cite{mehta2021modulated} to map a strand into a low-dimensional latent space, achieving superior reconstruction results.
In organizing the strands of a hair model, many previous works~\cite{zhou2018hairnet,rosu2022neural,lyu2022real} opt to parameterize the scalp area using UV unwrapping and assign strands to corresponding pixels.
While such a UV-mapping representation facilitates the exploitation of spatial adjacency among strands, due to the huge number of hair strands, a high-resolution UV map is often required, resulting in significant computational costs.
Taking advantage of the local similarity of human hairs, other works~\cite{chai2014reduced, chai2017adaptive, guan2012multi} propose to use a set of sparse guide strands as proxies for all hairs, leading to more efficient simulation, which is a common practice in the industry.
In this paper, we choose to utilize the strand-based representation for its fidelity and flexibility. We follow the convention of using guide hairs to represent hairstyles, which helps reduce the computational and memory overhead compared to representing individual strands directly.

\paragraph{Volumetric Representation of Hairs.}
An alternative approach to representing a hair model is through volumetric representation, where the entire hair volume is voxelized, and each voxel contains information about the growth direction and other properties of the hairs within it.
This voxelized representation often organizes the free-growing hairs into regular groups, making the hair structure easier to capture.
Although impressive results~\cite{saito20183d, yang2019dynamic, wu2022neuralhdhair, kuang2022deepmvshair, wang2022hvh} have been achieved, the expressiveness of volumetric representations is inherently limited by the granularity of the discretization.
The in-voxel fusion process inevitably leads to over-smoothed results, especially when dealing with curly hair or instances where hair strands cross each other within the same voxel.
Furthermore, the use of volumetric representation can be computationally expensive, particularly when dealing with large volumes of long hairstyles.

\paragraph{Hair Acquisition.}
Hair capturing is an active and evolving research area.
While a few previous works~\cite{herrera2012lighting, jakob2009capturing} seek to capture hair geometry with specialized devices, most existing methods for hair capturing rely on consumer-grade single- or multi-view cameras~\cite{paris2004capture, paris2008hair, wei2005modeling}.
Static hair reconstruction techniques~\cite{zheng2023hairstep, chai2012single, chai2013dynamic, luo2013structure, beeler2012coupled, nam2019strand, winberg2022facial, sun2021human, olszewski2020intuitive} aim to reconstruct 2D and 3D hair curves based on single-view visual cues and multi-view correspondences.
\revised{The work of~\cite{shen2020deepsketchhair} seeks to infer hair geometry from user sketches.}
In addition, dynamic capturing methods~\cite{xu2014dynamic, hu2017simulation, zhang2012simulation} take temporal consistency into consideration.
Although these model-free methods demonstrate impressive results, they tend to be fragile in challenging cases and unconstrained environments due to the thin strand geometry, occlusion between strands, and complex hairstyle configurations.
To enhance robustness, recent works have introduced prior models as constraints.
Some approaches~\cite{hu2014robust, chai2015high} employ geometric primitives as constraints for individual hairs during capturing, while others utilize hairstyle databases~\cite{hu2015single, chai2016autohair, liang2018video} for initialization and guidance, largely improving reconstruction quality and stability.
Furthermore, with the advancements in deep learning, neural approaches have emerged as the new state-of-the-art.
For instance, the work of~\cite{zhou2018hairnet} captures hair from monocular images by extracting hairstyle features using convolution neural networks.
The work of~\cite{rosu2022neural} combines multi-view reconstruction with neural descriptors to achieve photorealistic telepresence.
In addition to strand-based representations, volumetric methods~\cite{luo2012multi, saito20183d, yang2019dynamic, wu2022neuralhdhair, kuang2022deepmvshair} also have made significant progress in capturing both static and dynamic hair.
These works aim to provide reliable priors for robust and high-quality hair acquisition, presenting a potential application of our research.

\paragraph{Hair Generation.}
\revised{Compared to hair acquisition, generative hair models are relatively unexplored. A heuristic example-based method for hair generation is proposed by~\cite{ren2021hair}, which is largely limited by the reference hair models.}
Variational autoencoder (VAE) \cite{kingma2013auto} is widely recognized as one prevalent generative architecture.
In the context of hair modeling with volumetric representation, the work of~\cite{saito20183d} proposes the utilization of VAEs.
However, VAEs often suffer from over-smoothness issues despite their extensive examination for data embedding.
Generative adversarial networks (GAN)~\cite{goodfellow2014generative} have also demonstrated remarkable results in image generation~\cite{karras2019style, radford2015unsupervised}.
By incorporating the perceptual discriminator loss, GANs are particularly effective in recovering fine detail with weak supervision.
For our specific task, we employ both VAEs and GANs, where two VAE models encode individual strands and overall hairstyles, while another GAN model is adopted for detail restoration.

\section{Method}
\label{sec:method}

\begin{figure}[t]
\includegraphics[width=\linewidth]{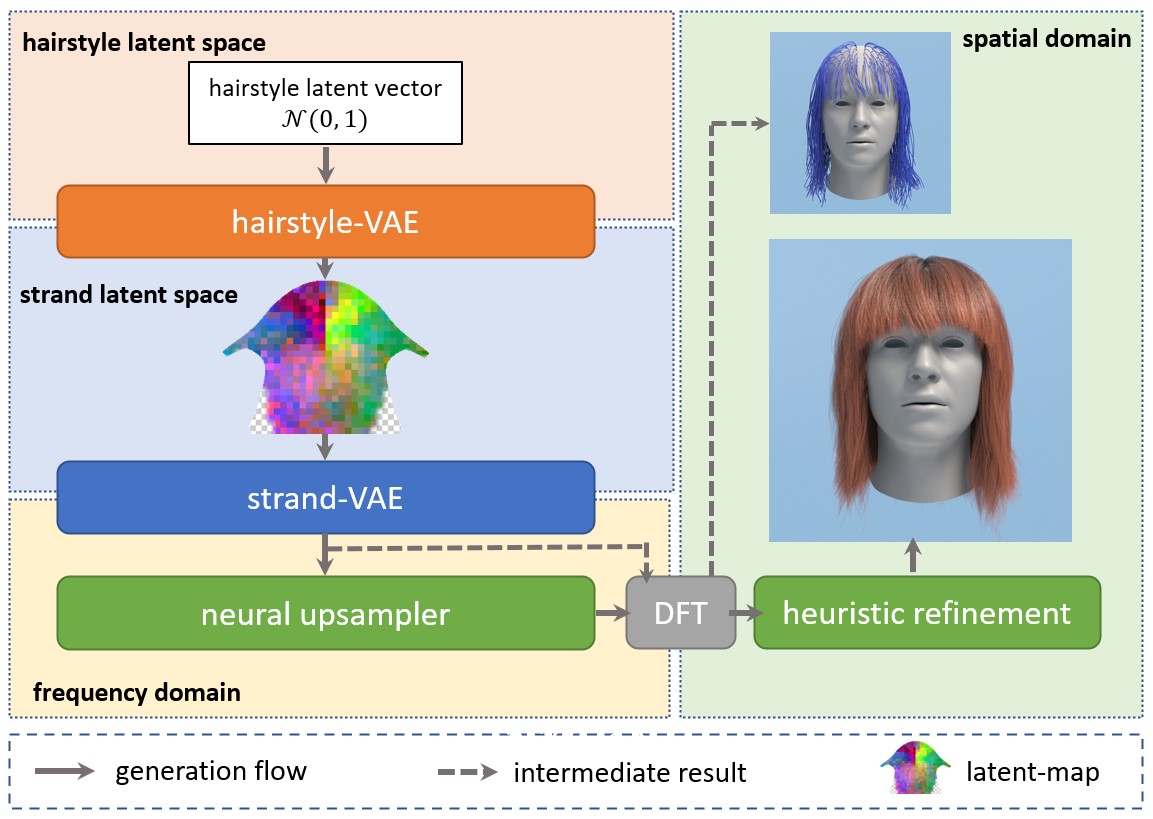}
\caption{
  We build a hierarchical representation of human hairstyles.
  To generate a hair model, we first draw a random vector from the Normal distribution in the hairstyle latent space.
  The vector is decoded by the hairstyle-VAE to get a low-resolution latent-map, which corresponds to sparse guide strands.
  Finally, the neural upsampler synthesizes dense hair strands from the sparse guide strands, which are further refined heuristically based on user specification.
}
\label{fig:pipeline}
\end{figure}

In this section, we present our comprehensive pipeline for hairstyle generation, tailored specifically for strand-based hair models.
Our algorithm operates on three hierarchical levels of human hair: individual strands, sparse guide strands, and the complete hair model with dense hair strands.
Accordingly, we design three components for each hierarchical level in hairstyle generation:
1) At the single strand level, our \textit{strand variational autoencoder} (\textit{strand-VAE}) establishes a low-dimensional latent space for encoding strands (\autoref{subsec:strand_vae}).
2) Building upon the strand latent space, a \textit{hairstyle variational autoencoder} (\textit{hairstyle-VAE}) further encodes a hair model, represented by a collection of sparsely-sampled guide strands, into a feature vector (\autoref{subsec:hairstyle_vae}).
3) A hybrid densification step consisting of a neural upsampler (\autoref{subsec:densifier}) and a heuristic refiner (\autoref{subsec:heuristic}) synthesizes the full hair style from the sparse guides.
By connecting all these components together, our pipeline provides a complete framework for hair model generation. The overall structure of the pipeline is illustrated in \autoref{fig:pipeline}.

\subsection{Strand Latent Space}
\label{subsec:strand_vae}

Our strand-VAE module performs the encoding of individual strands, transforming their Euclidean coordinates into a low-dimensional strand-wise latent space.
This latent space serves as the foundation for guide strand generation and quasi-static simulation.
Compared to raw Euclidean coordinates, our latent encoding is better regularized to a more constrained distribution of valid strands, without sacrificing strand-level geometric details, coherency among strands, or the overall diversity of the generated hairstyle.

Typically, a strand is represented as a polyline consisting of $N_s$ uniformly sampled points.
Previous work \cite{rosu2022neural} constructs a latent space based on this polyline representation.
However, we observe that representing strands in the original Euclidean space often has a significant negative impact on the structure preservation of the resulting latent space, leading to over-smoothed hair generation.
To better retain strand-level details like curliness, we parameterize the strands in the frequency domain using the discrete Fourier transform (DFT), and establish the strand latent space based on this frequency representation.
As elaborated in \autoref{subsec:ablation:svae}, this frequency latent space aligns better with human perception compared to the spatial latent space.

Formally, in the Euclidean spatial domain, a strand is originally represented as a polyline with $N_s$ points: $\mathcal{S}=\{\vp_1,\vp_2,\ldots,\vp_{N_s}\}\in\mathbb{R}^{N_s\times3}$ ($N_s=100$ in our case).
We first compute the gradients as
$\bar{\mathcal{S}}=\{\vd_1,\vd_2,\ldots,\allowbreak\vd_{N_s-1}\}\in\mathbb{R}^{(N_s-1)\times3}$
with the gradient displacement $\vd_i=\vp_{i+1}-\vp_i$,
and then equally divide the entire strand into $N_g$ non-overlapping segments:
$\bar{\mathcal{S}}_{i}=\{\vd_{i k},\vd_{i k+1},\allowbreak\ldots,\vd_{i k+k-1}\}, i \in \{1, 2, ..., N_g\}$
with $k=\lceil (N_s-1)/N_g \rceil$ denoting the segment size.
The strands are segmented to allow for varying shape statistics along the strand.
In all our experiments, we set $N_g=3$.
For each segment, we apply the DFT along $x, y, z$ axes separately with respect to vertex indices, obtaining coefficients of Fourier bases $\mathcal{F}_i^a \in \mathbb{C}^{f}$, with $a \in \{\mathrm{x}, \mathrm{y}, \mathrm{z}\}$ referring to the axes and $f = \lfloor k/2 \rfloor + 1$ representing the number of frequency bands.
Instead of using $\mathcal{F}$ directly, we further decompose it into three parts by taking their physical meanings into consideration:
$\mathcal{F}_A = \mathrm{abs}(\mathcal{F}) \in \mathbb{R}^{f}$,
$\mathcal{F}_\mathrm{cos} = \mathrm{real}(\frac{\mathcal{F}}{\mathcal{F}_A}) \in \mathbb{R}^{f}$,
and $\mathcal{F}_\mathrm{sin} = \mathrm{img}(\frac{\mathcal{F}}{\mathcal{F}_A}) \in \mathbb{R}^{f}$,
where $\mathrm{abs}(\cdot)$, $\mathrm{real}(\cdot)$, and $\mathrm{img}(\cdot)$ refer to the absolute value, real part, and imaginary part of a complex number.
Here, $\mathcal{F}_A$ describes the amplitude of each frequency, intuitively the significance of the strand's curliness and length;
$\mathcal{F}_\mathrm{cos}$ and $\mathcal{F}_\mathrm{sin}$ together describe the phase of the curves.
We encode the phase as vector $(\mathcal{F}_\mathrm{cos},\mathcal{F}_\mathrm{sin})$ instead of a scalar phase angle to avoid issues with periodicity.
Concatenating $\mathcal{F}_A$, $\mathcal{F}_\mathrm{cos}$, and $\mathcal{F}_\mathrm{sin}$ for all segments and axes, we obtain a vector $\mathcal{V}$, namely the \textit{frequency code}, of $N_g \times f \times 3 \times 3 = 459$ dimensions, to represent a strand in the frequency domain.

Our strand-VAE takes $\mathcal{V}$ as both the input and reconstruction target.
The employed training loss terms include: L1 loss $\mathcal{L}_{\mathrm{A}}$ for the amplitude coefficients $\mathcal{F}_\mathrm{A}$; L1 loss $\mathcal{L}_{\mathrm{P}}$ for the phase coefficients $\mathcal{F}_\mathrm{cos}$ and $\mathcal{F}_\mathrm{sin}$; and KL divergence loss $\mathcal{L}_{KL}^s$ with weight $\lambda_{KL}^s = 10^{-4}$:
\begin{equation}
  \mathcal{L}_{s} = \mathcal{L}_{\mathrm{A}} + \mathcal{L}_{\mathrm{P}} + \lambda_{KL}^s\mathcal{L}_{KL}^s.
\end{equation}
Since the phases of the high-amplitude components play a more crucial role, the phase loss $\mathcal{L}_{\mathrm{P}}$ on each frequency is weighted by the corresponding ground truth amplitude $\hat{\mathcal{F}}_\mathrm{A}$:
\begin{equation}
  \mathcal{L}_{\mathrm{P}} = \sum_{i=1}^{f} \bar{\mathcal{F}}_\mathrm{A}^i * (|\mathcal{F}^{i}_\mathrm{sin} - \hat{\mathcal{F}}^{i}_\mathrm{sin}| + |\mathcal{F}^{i}_\mathrm{cos} - \hat{\mathcal{F}}^{i}_\mathrm{cos}|),
\end{equation}
\begin{equation}
  \bar{\mathcal{F}}_\mathrm{A}^i = \frac{\hat{\mathcal{F}}_\mathrm{A}^i}{\sum_{j=1}^{f}\hat{\mathcal{F}}_\mathrm{A}^j}.
\end{equation}
Here, $\hat{\cdot}$ denotes the ground truth values and $\bar{\mathcal{F}}_\mathrm{A}^i$ represents the normalized weight for each frequency band $i$.
The summation over the axes and segments is omitted here for conciseness.

The encoder of the strand-VAE consists of a fully-connected network with $7$ layers.
Except the input and output layers, each layer has $1024$ hidden units with batch normalization \cite{ioffe2015batch} and residual connection \cite{he2016deep}.
It takes the individual strand representation $\mathcal{V}$ as input and compresses it into a latent code $\vl\in\mathbb{R}^{D_s}$.
We use $D_s = 64$ in our experiments, resulting in a compression rate of $21.5\%$.
On the other hand, the decoder follows the modulated sine network structure~\cite{mehta2021modulated} with $6$ layers and $1024$ hidden units.
Given a latent code $\vl$, the decoder generates the vector $\mathcal{V}$, which can then be converted back to Euclidean coordinates $\mathcal{S}$, with the additional input of the root position $\vp_1$ pre-defined on the scalp.

\subsection{Scalp Space Hairstyle Parameterization}
\label{subsec:scalp_space}

\begin{figure}[t]
\includegraphics[width=\linewidth]{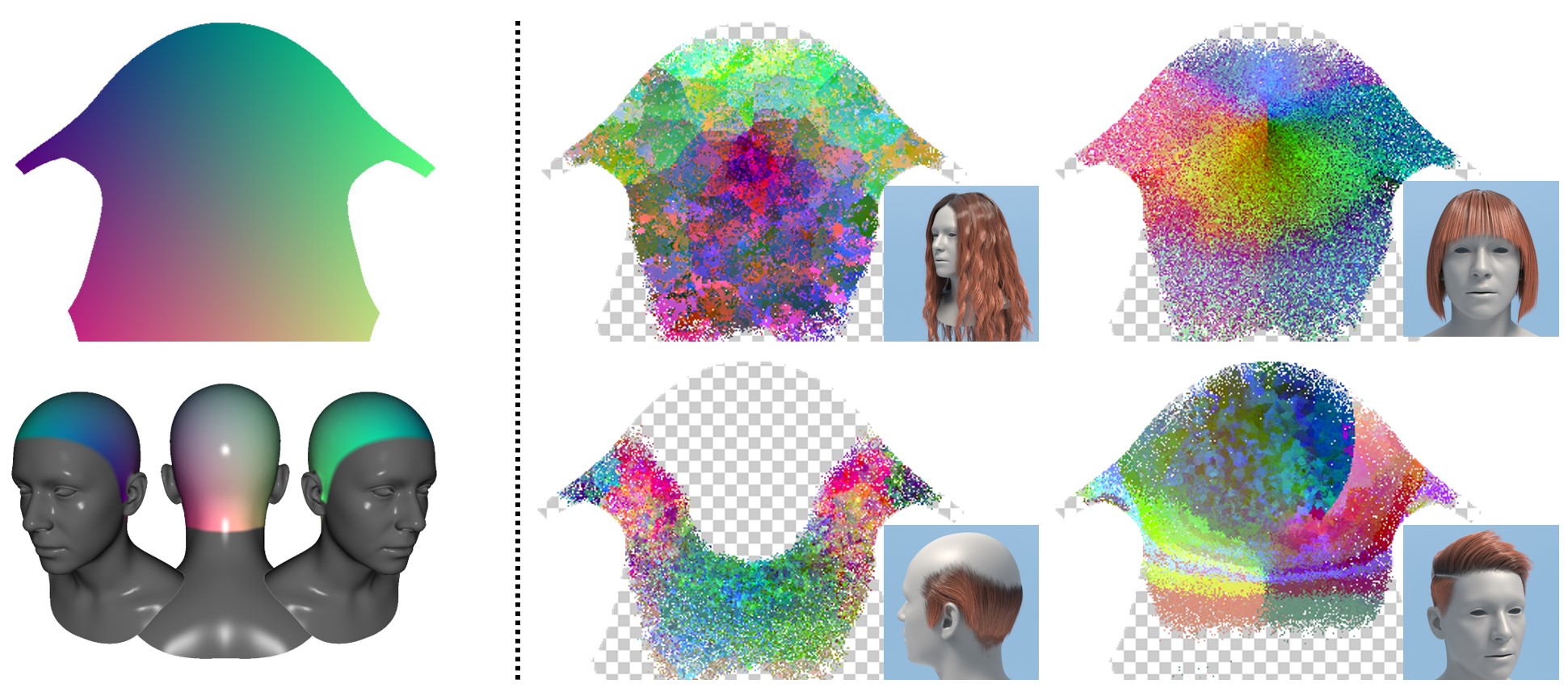}
\caption{
  Illustration of our scalp parameterization (left) and latent-maps with baldness (right).
  Visualized using selected 3 axes of the strand latent codes.
}
\label{fig:scalp_space}
\end{figure}

We now introduce our representation of hairstyles.
Similar to prior works~\cite{wang2009example,rosu2022neural,lyu2022real,zhou2018hairnet}, we define the 2D parameterization of hairs on the scalp surface as a regular UV map, as illustrated in Fig.~\ref{fig:scalp_space}.
The strand representations are embedded into the UV map at the positions corresponding to their roots on the scalp.
\revised{When the strands are represented by frequency codes $\mathcal{V}$, the corresponding UV map is referred to as a \textit{strand-map}; when the strands are represented by their latent codes, the UV map is called a \textit{latent-map}. These two maps can be mutually converted using the strand-VAE.}

In our pipeline, we employ two different resolutions for the latent-maps: $24\times32$ (referred to as a low-resolution map where $1$ pixel has side length $1.0-2.9$cm) and $216\times288$ (referred to as a high-resolution map where $1$ pixel has side length $0.1-0.3$cm).
\revised{As not all texels are used, the low-resolution maps usually accommodate around 300 hairs, while the high-resolution ones contain around 25K hairs.}
Initially, we generate a low-resolution latent-map that corresponds to sparse guide strands, and then adopt a hybrid densification step to generate dense hair strands from the guide strands.
This design choice is motivated by the observation of high redundancy in dense hairs due to the local coherency of nearby strands.
Compared to directly using high-resolution latent-maps ($256\times256$ in \cite{rosu2022neural} and $128\times128$ in \cite{lyu2022real}), our intermediate representation enables better convergence during training and higher computational efficiency during inference.
This design also aligns with the common CG practice of using sparse guide strands to model and control the global hairstyle structure before densification.

Additionally, to ensure generalizability to a broader range of hairstyles, we incorporate baldness as part of the hairstyle.
Baldness is defined by an additional binary mask, referred to as a \textit{baldness-map} within the scalp space, as depicted in Fig.~\ref{fig:scalp_space}.

\subsection{Hairstyle Latent Space}
\label{subsec:hairstyle_vae}

Based on the scalp space parameterization, our hairstyle-VAE learns to generate whole hairstyles utilizing the VAE framework.
The input and reconstruction target for the hairstyle-VAE consist of both the low-resolution latent-map $\mathcal{M}_l \in \mathbb{R}^{w_M \times h_M \times D_s}$ and the baldness-map $\mathcal{M}_b \in \mathbb{R}^{w_M \times h_M}$, where $w_M$ and $h_M$ represent the width and height of both maps.
Within the hairstyle-VAE, the encoder projects the latent-map $\mathcal{M}=\{\mathcal{M}_l,\mathcal{M}_b\}$ into a single latent vector $\vh\in\mathbb{R}^{D_h}$ (we set $D_h=512$, resulting in a compression rate of $99\%$), and the decoder takes the latent vector $\vh$ as input to reconstruct the corresponding latent-map.
The training objective is defined as:
\begin{equation}
\mathcal{L}_{h} = \mathcal{L}_\mathrm{rec}^{\mathcal{M}_l} + \mathcal{L}_\mathrm{rec}^{\mathcal{M}_b} + \lambda_{KL}^h\mathcal{L}_{KL}^h,
\end{equation}
where $\mathcal{L}_\mathrm{rec}^{\mathcal{M}_l}$ and $\mathcal{L}_\mathrm{rec}^{\mathcal{M}_b}$ are the L1 reconstruction losses for latent-map $\mathcal{M}_l$ and baldness-map $\mathcal{M}_b$, and $\mathcal{L}_{KL}^h$ is the KL divergence loss with weight $\lambda_{KL}^h = 0.01$.

Our hairstyle-VAE utilizes a concise network architecture.
The encoder part consists of a total of $12$ convolutional layers, incorporating residual connections.
Similarly, the decoder is symmetric to the encoder and employs transposed convolutions for upsampling.
Please see \autoref{sec:network-detail} for detailed network structures.

\subsection{Neural Upsampling}
\label{subsec:densifier}

The hairstyle-VAE produces a low-resolution latent-map that represents sparse guide strands.
To further generate a complete hair model with around $150$K strands, we then employ a hybrid densification process involving two steps: upsampling and refinement.
\revised{The upsampling step outputs a high-resolution strand-map with 25K hairs, and the refinement step additionally populates the strands by 6 times.}

We emphasize that an end-to-end model is less suitable here because the mapping from sparse guide strands to dense hair strands is one-to-many, and the user's involvement is often necessary to resolve the ambiguity.
Our two-step hybrid approach strikes a balance between simplicity and controllability.
In the first step, our novel neural upsampler automatically populates the strands based on the guide strands.
In the second step, users are allowed to refine the high-frequency details, providing control over the final results.
In this section, we will introduce the first step of neural upsampling, while the second step will be elaborated in \autoref{subsec:heuristic}.

\begin{figure}[t]
  \includegraphics[width=\linewidth]{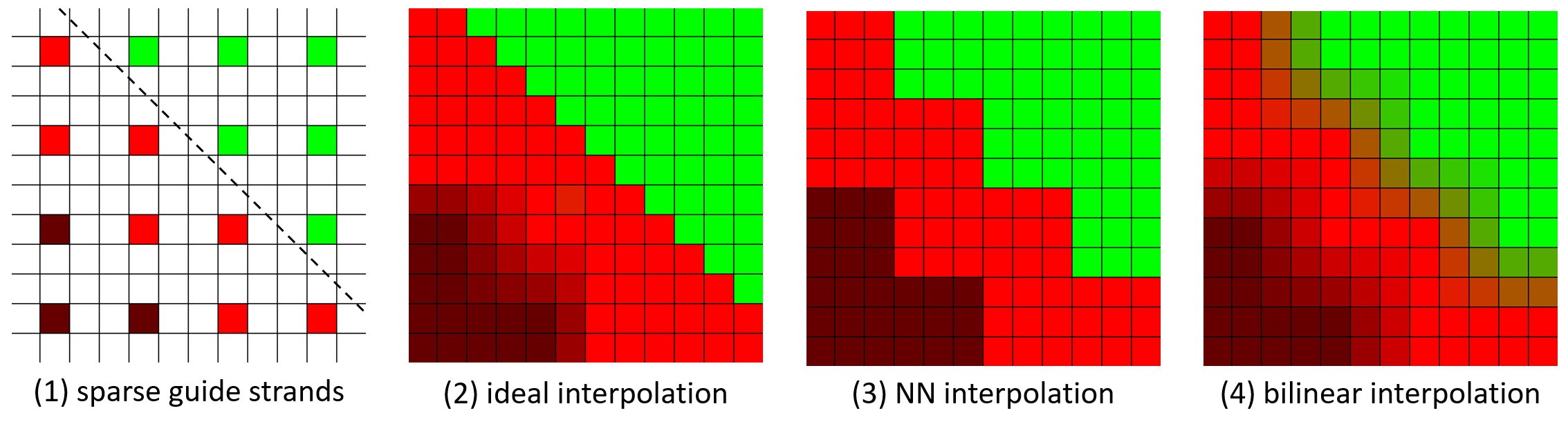}
  \caption{
    Illustration of interpolating near a parting line.
    (1) shows the source low-resolution strand-map where red and green represent strands growing in opposite directions and the dashed line is the expected parting line.
    The ideal interpolation is (2) where the parting line is sharp while other regions are smooth.
    The nearest-neighbor interpolation in (3) has aliasing artifacts.
    The bilinear interpolation in (4) smooths out the parting line and the strands may even penetrate the head mesh.
  }
  \label{fig:interp_illus}
\end{figure}

In the upsampling step, we aim to estimate a high-resolution strand-map from a low-resolution one where pixels contain frequency representations $\mathcal{F}$ of guide strands.
The low-resolution strand-map is obtained by decoding the output of the hairstyle-VAE with the strand-VAE.
Analogous to image upsampling, we assume that each dense strand can be viewed as a linear interpolation of its four neighboring guide strands.
Consequently, the task simplifies to estimating the interpolation weights of the guide strands at each position.
Trivial interpolation methods are inadequate for this task due to the significant variation in local smoothness observed in human hairstyles.
Bilinear interpolation, for example, smooths out sharp parting lines and result in hair-head penetration, while nearest-neighbor interpolation exhibits aliasing artifacts (as depicted in \autoref{fig:interp_illus}).
Therefore, spatially varying interpolation is often preferred, but determining an effective interpolation strategy \textit{a priori} is challenging.
While the industry relies on intensive manual design for this purpose, we introduce a \textit{neural upsampler} to automate the process and ensure realism.

The input of the neural upsampler is a high-resolution multi-channel feature map that describes the guide strands distribution.
At each pixel, the feature vector is formed by concatenating the low-frequency components of its four neighboring guide strands, their bilinear interpolation, and the distances to the guides.
Only the low-frequency components are considered, with a cut-off frequency set to $f_l = 8$.
This choice is made because the high-frequency details have less significance in the interpolation process and will be refined later on.
The output of the neural upsampler is a $5$-channel \textit{weight-map}, where the first $4$ channels represent the weights for the four neighboring guide strands, and the last channel represents the weight for the bilinear interpolation of the guide strands.
\revised{At inference, each strand $Y$ on the interpolated high-resolution map is computed as $Y=a_1X_1+a_2X_2+a_3X_3+a_4X_4+a_5 \mathcal{B}(X_1,X_2,X_3,X_4)$, where $\{a_i\}$ represent the predicted weights, $\{X_i\}$ represent the neighboring guide hairs from the low-resolution map, and $\mathcal{B}(\cdot)$ denotes the bilinear interpolation of guide hairs at the position of $Y$.
The bilinear interpolation is an effective shortcut because oftentimes it already provides a solution close to the optimal one.}

The neural upsampler is trained in an adversarial framework (GAN) \cite{goodfellow2014generative} for two reasons:
First, dense hair interpolation is not a deterministic task and there is no unique ground truth;
Second, for dense hairstyles, it is more reasonable to perceptually evaluate the hair model as a whole rather than enforcing per-strand supervision.
We devise the neural upsampler as a $12$-layer convolutional network with residual connections and instance normalization \cite{ulyanov2016instance}, and use a large kernel size of $13$ to perceive more spatial information in this high-resolution setting (more details in \autoref{sec:network-detail}).
The discriminator has the same structure as the generator, except that its input is the interpolated high-resolution strand-map and its output is a score map.
The loss function follows the Wasserstein loss \cite{arjovsky2017wasserstein}.
Denoting the neural upsampler as $G$ and the discriminator as $D$, the loss function for the discriminator is:
\begin{equation}
  \mathcal{L}_{D} = D(\mathcal{H}) - D(\mathcal{X}) + D(\mathcal{H})^2 + D(\mathcal{X})^2,
\end{equation}
where $\mathcal{X}$ is a real strand-map from the dataset and $\mathcal{H}$ is a generated one.
$D(\mathcal{H})^2$ and $D(\mathcal{X})^2$ are regularization terms that prevent $D(\mathcal{H})$ and $D(\mathcal{X})$ from being numerically too large.
The loss function for the generator is:
\begin{equation}
  \mathcal{L}_{G} = - D(\mathcal{H}) + \lambda_{G} (\mathcal{L}_{G}^{\mathrm{bl}} + \mathcal{L}_{G}^{\mathrm{g}} + \mathcal{L}_{G}^{\mathrm{sum}}).
\end{equation}
Denoting the $5$-channel interpolation weights estimated by $G$ as $\mathcal{W}$, $\mathcal{L}_{G}^{\mathrm{bl}} = |\mathcal{W}_{5} - 1|$ biases the weight of bilinear interpolation (channel $5$) towards 1, $\mathcal{L}_{G}^{\mathrm{g}} = \sum_{i=1}^{4} |\mathcal{W}_{i}|$ regularizes the weights of each guide towards $0$, and $\mathcal{L}_{G}^{\mathrm{sum}} = |\sum_{i=1}^{5} \mathcal{W}_{i} - 1|$ softly normalizes the weights.
The regularization weight $\lambda_{G}$ is set to $0.1$.

\subsection{Heuristic Refinement}
\label{subsec:heuristic}
\revised{The high-resolution map generated by the neural upsampler contains 25K strands, which is still fewer than normal human hairs.}
To further enhance the quantity and quality of the dense strands, we introduce a heuristic refinement step that increases the number of hairs to 150K with fine details.
In this step, we provide the user with creative semantic control over the final appearance.
This step can also be fully automatic with fixed or randomized parameters if a hands-off approach is preferred, e.g. for large scale data generation.

We start by addressing penetrations (detailed in \autoref{sec:penetration}) and perturbing the \textit{frequency representation} $\mathcal{F}$ with random noise to increase variation.
The scale of the noise can be specified by the user, allowing for the creation of regular or messy hairstyles.
Next, we perform wisp formation in the Euclidean spatial domain.
The user may specify two parameters: the number of wisps $w$ and the stickiness $s$ to control the clustering of hairs.
We adopt k-means clustering to identify $w$ wisp clusters, and then guide each strand towards the center of its corresponding cluster:
\begin{equation}
  \delta_i = \frac{s \cdot \mathrm{min}(1, \frac{l_i}{\bar{l}})}{\mathrm{max}(1, d^2_i)} + \sum_{k=1}^{i-1}\delta_k.
\end{equation}
We denote the vertex index as $i$, where $i = 1$ is the fixed strand root with displacement $\delta_1 = \mathbf{0}$, and the deformation $\delta_i$ of other vertices is determined by stickiness $s$, distance to the center strand $d_i$, and length to the strand root $l_i$.
To prevent excessive deformation near the root, we empirically set $\bar{l}$ to $5$cm as a threshold.
This simple deformation strategy can yield practically satisfactory results since the neural upsampler provides a good initialization.
This wisp formation is skipped when $w = 0$ or $s = 0$.
Finally, as the raw output of the neural upsampler only contains $25K$ strands, we duplicate all strands $6$ times with small variations in the frequency domain again.
This allows us to increase the total strand number of the final model to $150K$, further enhancing its density and realism.

\section{Experiments and Applications}
\label{sec:exp}

Extensive experiments are conducted to validate the effectiveness of our hair generation pipeline.
In \autoref{subsec:dataset}, we introduce the dataset, training procedure, and system runtime.
In \autoref{subsec:evaluation}, we evaluate each main component of our method.
We present ablation studies to justify the major technical choices in \autoref{subsec:ablation}.
\revised{Finally, in \autoref{subsec:simulator}, we introduce a quasi-static neural hair simulator as one downstream application of our model.}

\subsection{Datasets, Training, and Runtime}
\label{subsec:dataset}

The model is trained and evaluated on an artist-created hairstyle dataset, referred to as \textsc{GroomHair}, which comprises diverse hairstyles with fine-grained variations.
To create the dataset, the artists first identified $35$ base hairstyle categories, encompassing a wide range of styles such as buzz, bobby, pixie, wavy, afro, and more \revised{(see \autoref{sec:dataset-list} for the complete list)}.
For each category, the artists utilize \textit{Houdini}~\footnote{https://www.sidefx.com/products/houdini/} to create a recipe that defines the desired hairstyles and generate a series of fine-grained variations ($50-400$ depending on the hairstyle) of the same category using different parameters.
In some categories, baldness is also modeled, which corresponds to the baldness-map used by the hairstyle-VAE.
The final dataset contains $7712$ data samples, each representing a specific hair model with approximately $150$K strands.
We randomly split the entire dataset into $6940$ training samples and $772$ test samples.
The training samples are further augmented by horizontal mirroring.

We first train our strand-VAE model using the \textsc{GroomHair} dataset.
Subsequently, we fix the strand-VAE model and use it to process the hair models in \textsc{GroomHair} to obtain the training and testing data for the hairstyle-VAE.

The strand-VAE and hairstyle-VAE are both trained using the Adam~\cite{kingma2014adam} optimizer with an initial learning rate of $10^{-3}$, which is reduced by a factor of $0.1$ whenever the training loss ceases to improve.
The training process continues until the learning rate reaches $10^{-6}$.
The neural upsampler is also trained using the Adam optimizer but with a fixed learning rate of $10^{-4}$ for a total of $105$K iterations (around $27$ epochs).

We evaluate the runtime performance of each module on a PC equipped with an Intel Core i9-11900KF CPU and an NVIDIA A100 40GB GPU. The results are summarized in \autoref{tab:runtime}.
It is noteworthy that the runtime measurements for the strand-VAE and hairstyle-VAE only include the decoder parts of the networks.
Our system exhibits real-time performance, achieving more than $30$ FPS for the generation and simulation of $300-500$ guide strands, which is a common quantity in hairstyle authoring.
Users can edit the hairstyle by either tweaking the hairstyle code in the latent space or modifying the guide strands in the Euclidean space interactively.
The heuristic refinement step typically takes $10-15$ seconds, and the optional penetration correction step takes $30-55$ seconds.

\begin{table}[t]
  \caption{
    Runtime and number of parameters of the modules.
    Our system achieves real-time performance for generation, editing, and simulating up to 500 strands before densification.
  }
  \label{tab:runtime}
  \begin{tabular}{ccccccc}
    \multicolumn{5}{c}{component runtime (ms)} & & \multirow{2}{*}{\# params.} \\
    \cline{1-5}
    \# strands & 1 & 300 & 500 & 150K & & \\
    \hline
    \hline
    strand-VAE & 1.90 & 14.9 & 23.5 & 7670 & & 10.59M \\
    \hline
    neural simulator & 3.26 & 4.58 & 4.77 & 1330 & & 32.74M \\
    \hline
    \hline
    hairstyle-VAE & \multicolumn{4}{c}{3.42} & & 83.65M \\
    \hline
    neural upsampler & \multicolumn{4}{c}{281} & & 11.22M \\
    \hline
  \end{tabular}
\end{table}

\subsection{Evaluation}
\label{subsec:evaluation}

In this section, we provide a comprehensive assessment of the components of our system in a bottom-up order.
First, we evaluate the performance of the strand-VAE in encoding individual strands into the strand latent space.
Next, we assess the hairstyle-VAE in embedding a given hair model into the hairstyle latent space, as well as generating hair models from this latent space.
Finally, we demonstrate how the densification step effectively produces realistic dense hairs from sparse guide strands.

We use the following per-strand metrics.
Recall that originally each hair strand is represented as a polyline $\mathcal{S}=\{\vp_1,\vp_2,\ldots,\vp_{N_s}\}$ and the parent-relative displacement is defined as $\vd_i = \vp_{i+1} - \vp_i$.
Positional error (pos. err.) calculates the mean distance between corresponding points of the predicted strands and the ground truth (indicated by $\hat{\cdot}$): $\sum_i^{N_s}\|\vp_i-\hat{\vp_i}\|/N_s$.
Local position error (loc. err.) measures the L2 distance between the gradients of corresponding points on the strands, without accumulating errors along the hair: $\sum_i^{N_s-1}\|\vd_i-\hat{\vd_i}\|/(N_s-1)$.
We report these metrics by averaging them per hair model and then across the entire test set. This ensures that each hair model contributes equally to the final numbers.

\paragraph{Strand-VAE}

\begin{table}[t]
  \caption{
    Quantitative metrics of the strand-VAE, hairstyle-VAE, and neural simulator.
    The errors are acceptably low.
  }
  \label{tab:eval-quant}
  \begin{tabular}{cccc}
    \hline
     & strand-VAE & hairstyle-VAE & neural simulator \\
    \hline
    pos. err. & 1.90mm & 7.26mm & 8.89mm \\
    \hline
    loc. err. & 0.15mm & 0.40mm & 0.44mm \\
    \hline
  \end{tabular}
\end{table}

\begin{figure}[t]
  \includegraphics[width=\linewidth]{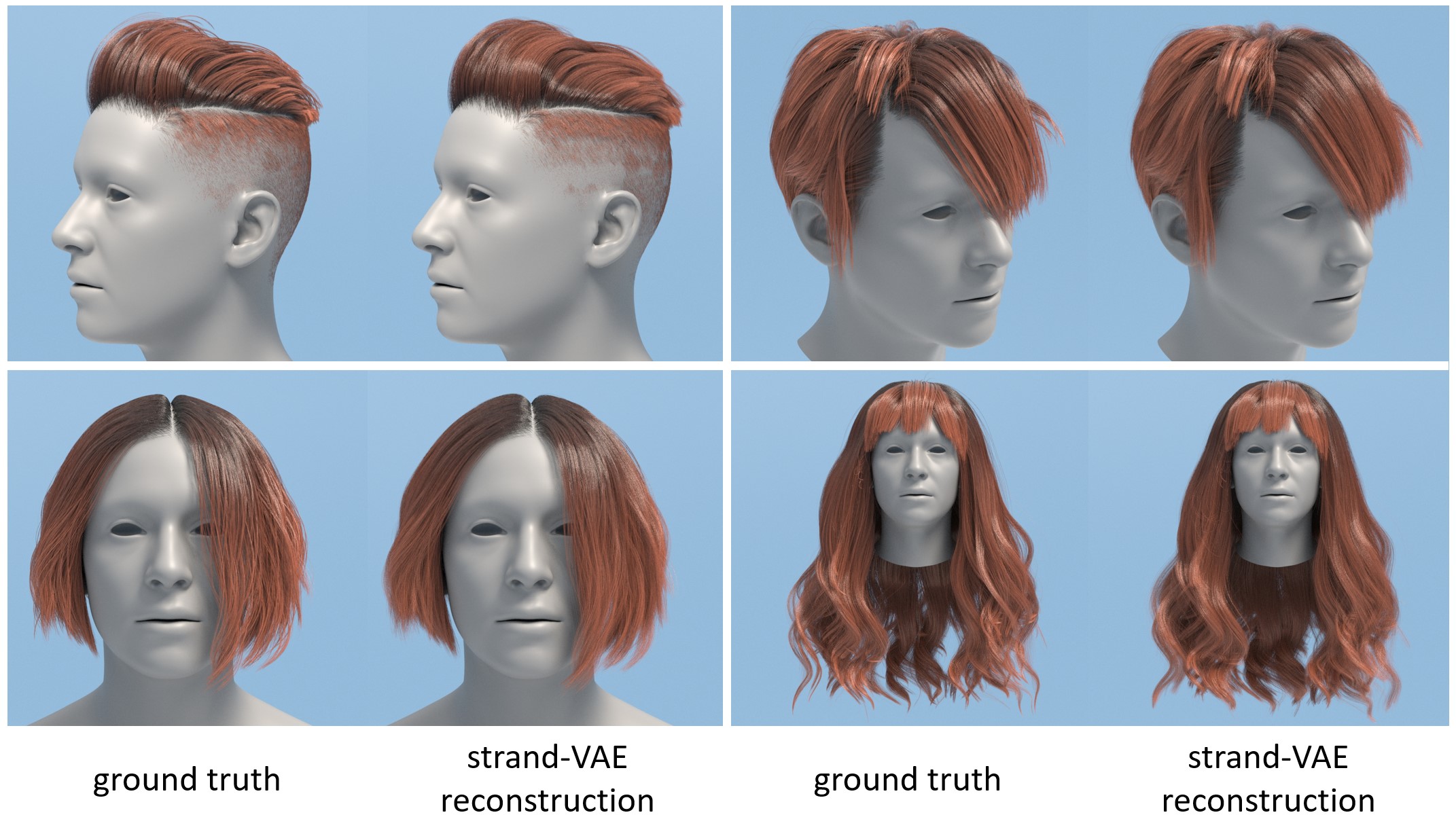}
  \caption{
    Reconstruction results of strand-VAE on the test set by encoding and decoding each strand.
    The difference is barely observable.
    The average positional errors for the demonstrated samples are (left to right, top to bottom): 1.41mm, 1.13mm, 1.30mm, and 5.08mm.
  }
  \label{fig:sae-test}
\end{figure}

We first evaluate the strand-VAE model on the test set by measuring the reconstruction error of encoding and decoding individual strands.
The quantitative results are reported in \autoref{tab:eval-quant} (1st column).
The mean reconstruction error is remarkably low, measuring only $1.90$mm.
In \autoref{fig:sae-test} we provide a few examples of strand reconstruction from the test set.
As demonstrated, the reconstruction is of high fidelity and the difference is hard to discern.
These results indicate that the strand-VAE effectively constructs a latent space that serves as a solid foundation for subsequent steps.

\paragraph{Hairstyle-VAE}

\begin{figure}[t]
  \includegraphics[width=\linewidth]{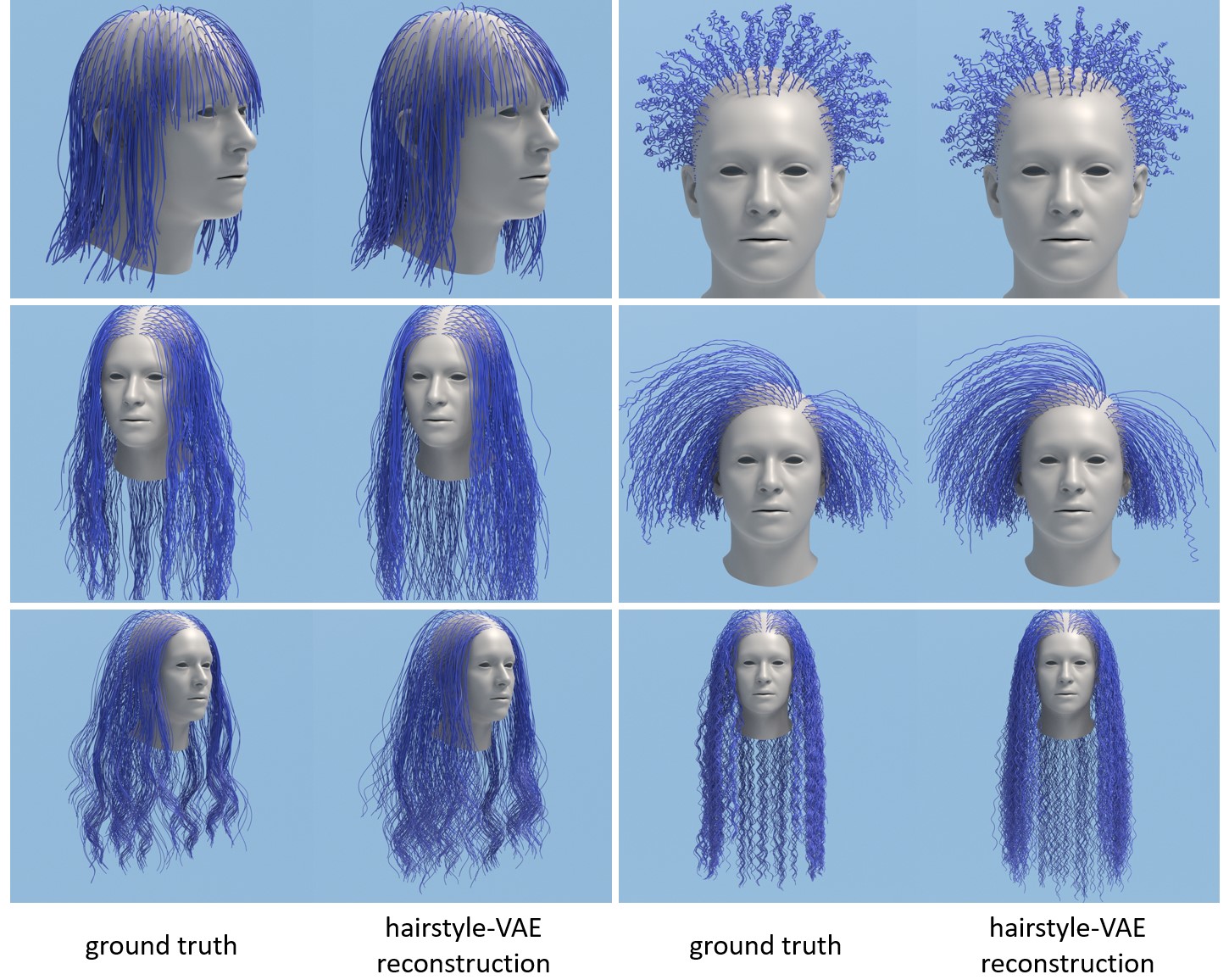}
  \caption{
    Selected results of our hairstyle-VAE on the test set.
    Hairstyles are well-preserved with only 0.4\% parameters of the original guide hairs.
    The average positional error for the demonstrated samples are (left to right, top to bottom):
    8.35mm, 5.65mm, 24.53mm, 12.76mm, 24.61mm, and 20.68mm.
  }
  \label{fig:hae-test}
\end{figure}

\begin{figure*}[t]
  \includegraphics[width=\linewidth]{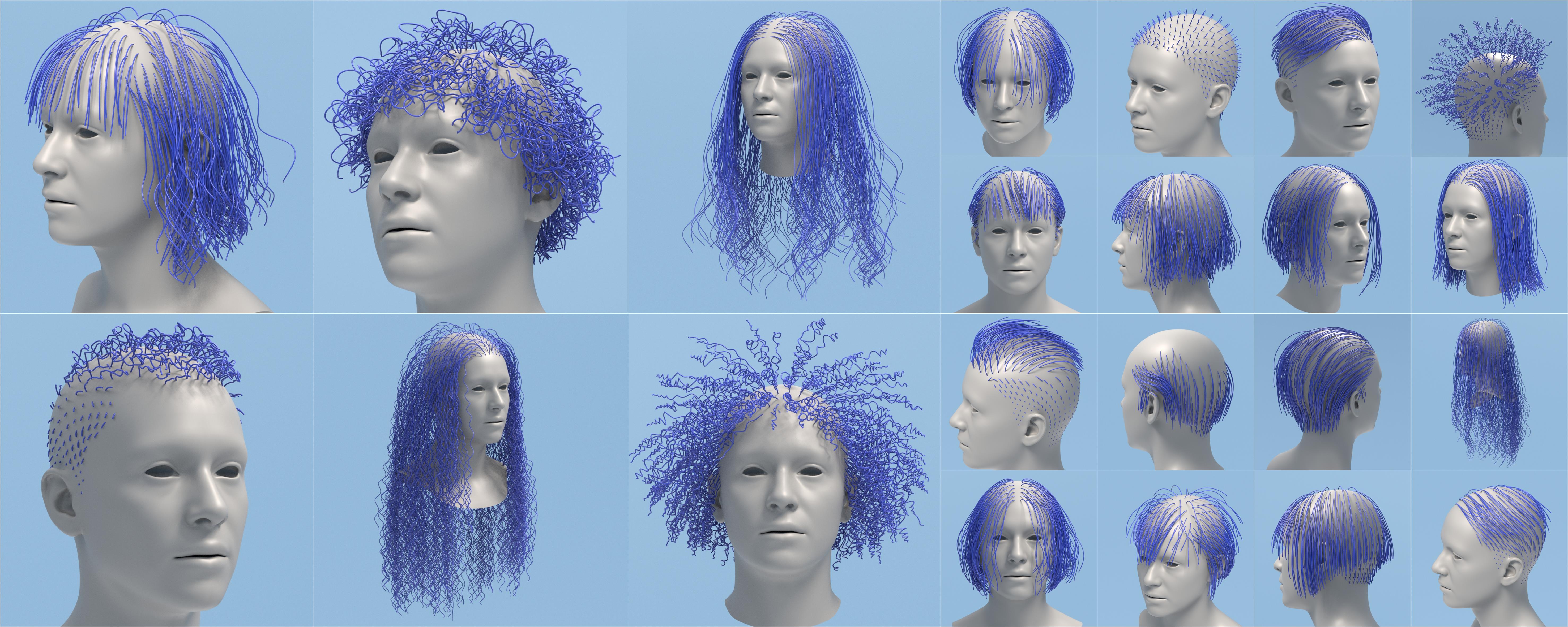}
  \caption{
    Diverse hairstyles generated by our hairstyle-VAE from random vectors sampled in the hairstyle latent space.
    We would like to emphasize
    1) the diversity of hairstyles, which comes from the powerful hairstyle-VAE;
    2) high local variety of the hairs, which originates from the well-structured frequency strand latent space.
  }
  \label{fig:hae-random}
\end{figure*}

\begin{figure}[t]
  \includegraphics[width=\linewidth]{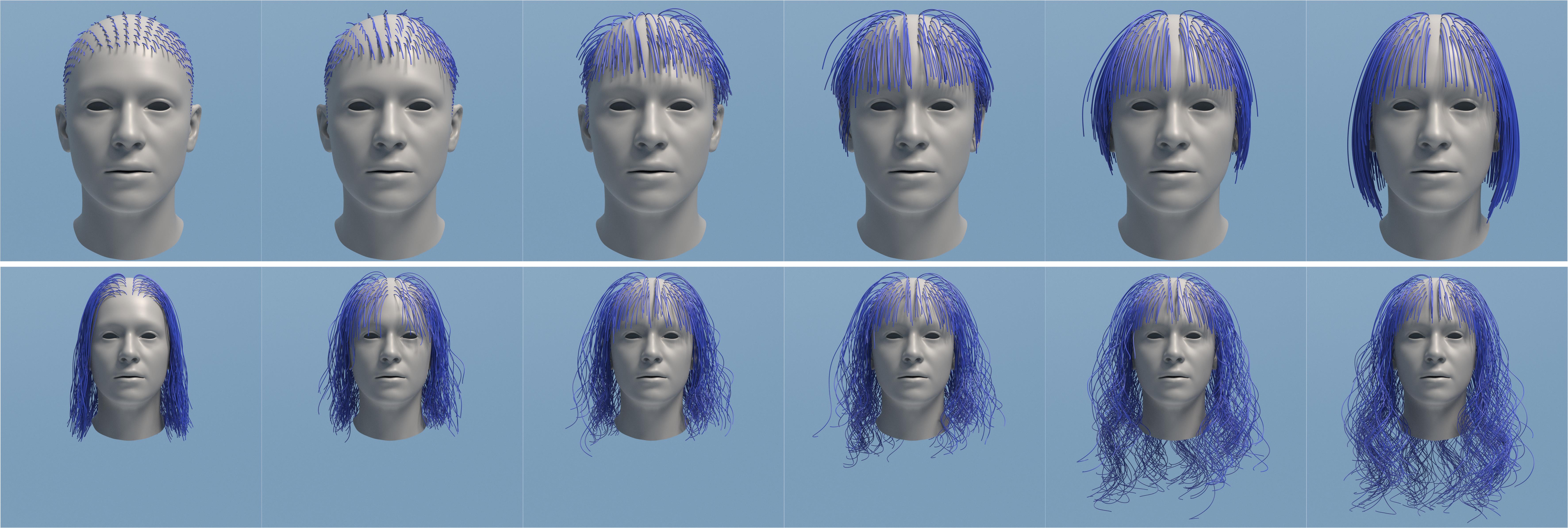}
  \caption{
    Interpolation of hairstyles from left to right in the hairstyle latent space.
    While the start and end hairstyles are distinct, the interpolation trajectory is reasonable.
  }
  \label{fig:hae-interp}
\end{figure}

\begin{figure}[t]
  \includegraphics[width=\linewidth]{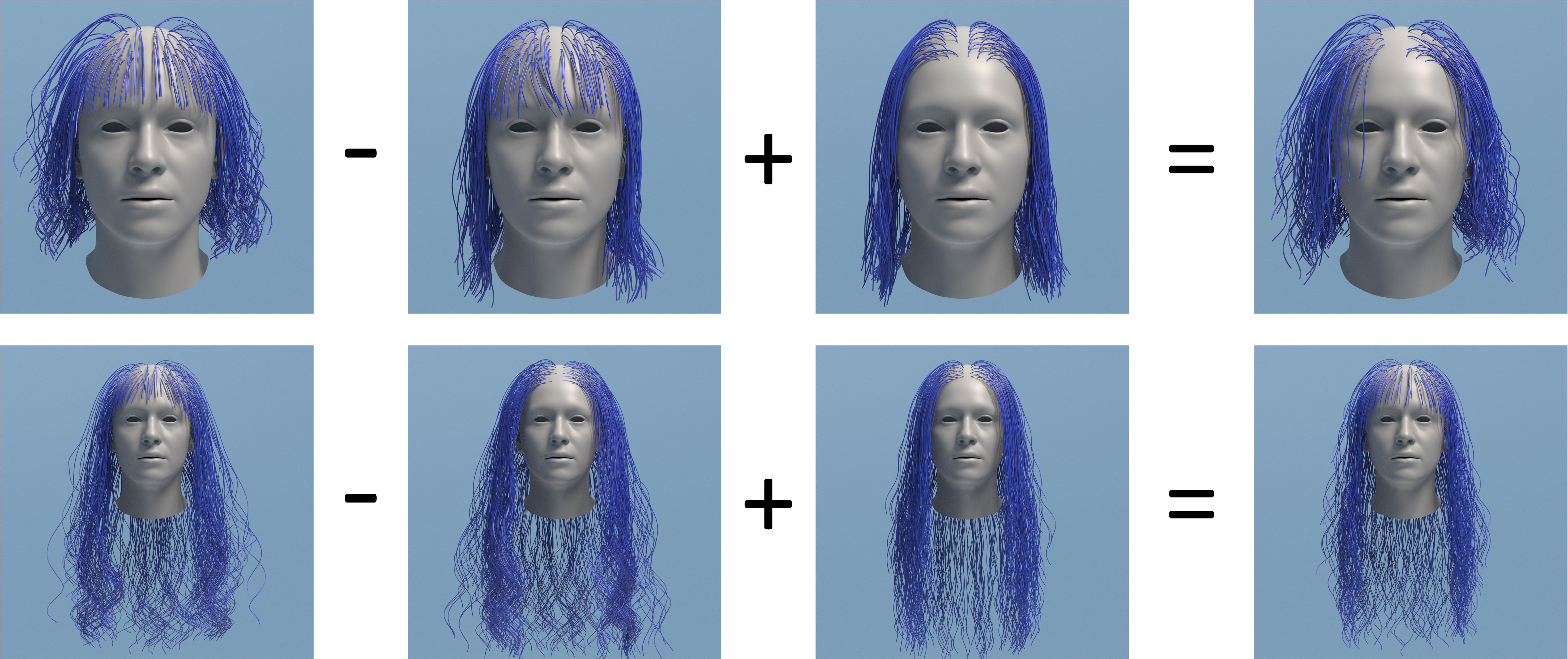}
  \caption{
    Arithmetic between hairstyles: we subtract the hairstyle latent vector $\vh$ of column 2 from column 1 and add the difference to column 3, finally get column 4.
    First row: the differences in curliness and length are successfully transferred.
    Second row: the difference between column 1 and 2 is the fringe, which is added to column 3 and gives us column 4.
    The hairstyles in the last column are \textit{not} in the original dataset.
  }
  \label{fig:hae-arith}
\end{figure}

Next, We evaluate the performance of the hairstyle-VAE model by encoding and decoding the entire hair model represented as a latent-map.
The reconstructed latent-map is then decoded by the strand-VAE to obtain hair strands in the Euclidean space for visualization and error computation.
The resulting reconstruction errors are reported in \autoref{tab:eval-quant} (2nd column).
As encoding the entire hair model is generally more challenging, particularly in our setting where the hairstyle latent vector utilizes only $0.4\%$ parameters of the guide hairs, we observe relatively larger errors compared to the strand-VAE.
Nevertheless, the average error of $7$mm remains at an acceptably low level, preserving the overall visual appearance of the hairstyle well as can be seen in \autoref{fig:hae-test}, even for challenging hairstyles (right column) that have not been explored in previous works.
Additionally, our model accurately reconstructs the baldness-map, a component often overlooked in prior work.
The intersection-over-union (IoU) of the baldness-map on the test set measures $97.3\%$ when the threshold is set to 0.8.

We now present a series of experiments to showcase the capability of the hairstyle-VAE in hairstyle generation and authoring.
Firstly, we demonstrate that the latent space of the hairstyle-VAE is sufficiently well-constrained to allow for meaningful hairstyle generation by direct random sampling from the Normal distribution, as shown in \autoref{fig:hae-random}.
It is worth emphasizing the high local diversity observed in the generated hair styles, where the strands deviate from each other frequently, in contrast to previous works where nearby strands tend to grow in parallel and become over-smoothed.

Additionally, we show that hairstyles can also be generated by traversing the latent space through interpolation, as shown in \autoref{fig:hae-interp}.
Despite the distinct characteristics of the starting and ending hair models, the interpolation path in the hairstyle latent space remains semantically valid. This demonstrates the flexibility of our model in generating new hairstyles with controllable attributes.

Furthermore, we explore another interesting application of arithmetic operations between hairstyles, as shown in \autoref{fig:hae-arith}.
Denoting the hairstyle latent vector of column $i$ as $\vh_i$, we compute $\vh^* = \vh_1 - \vh_2 + \vh_3$ and decode $\vh^*$ to produce the resulting hair model.
This simple arithmetic aligns well with human intuition and can generate novel hairstyles that do not exist in our dataset.

The most similar previous work is Volumetric Hair VAE (VHV) of \cite{saito20183d} that learns the latent embedding of hair models from volumetric representation.
However, a fair comparison is challenging due to fundamental differences in tasks (capture vs. generation), hair representations (voxels vs. strands), and datasets.
For a rough comparison, we convert our predicted strands into volumetric representations and report the numbers in \autoref{tab:comp-usc} (note that the test sets differ).
The metrics of IoU, precision, and recall evaluate the correctness of strand occupancy in the space.
While our latent space is much more compact than that in VHV, our method still outperforms VHV on these metrics.
It is noteworthy that our method exhibits higher error on the growing flows due to the fact that the volumetric representation in VHV fuses growing directions within the same voxel, which conflicts with our strand representation without any fusion.
Furthermore, our model can generate hairstyles simply from random sampling, while VHV only demonstrates interpolation of given similar hairstyles.
\revised{Please find qualitative comparisons with VHV in \autoref{sec:vhv-comp}.}

\begin{table}[t]
  \caption{
    A \textbf{rough} comparison with VHV \cite{saito20183d}.
    Our method has higher IoU, precision, and recall, which suggests better estimation of strand occupancy in the space.
    Our method has a slightly higher L2 flow error because we use a strand-based representation.
    Notably, ours has a much more compact latent space that supports direct random sampling.
  }
  \label{tab:comp-usc}
  \begin{tabular}{cccccc}
    \hline
     & IoU & Precision & Recall & L2 (flow) & latent dim. \\
    \hline
    VHV & 0.8243 & 0.8888 & 0.9191 & 0.2118mm & 6144 \\
    \hline
    ours & 0.9426 & 0.9777 & 0.9626 & 0.2807mm & 512 \\
    \hline
  \end{tabular}
\end{table}

\paragraph{Hybrid Densification}
\label{subsec:hybrid}

\begin{figure*}[t]
  \includegraphics[width=\linewidth]{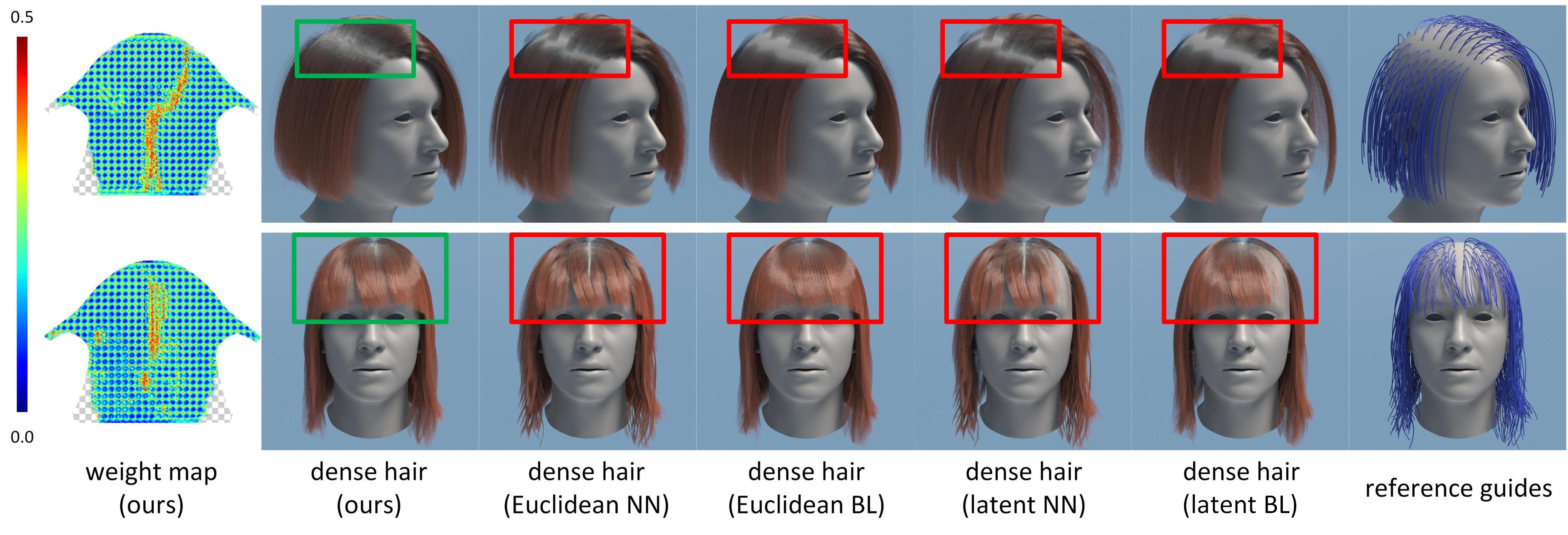}
  \caption{
    Output from our neural upsampler based on our hairstyle-VAE reconstruction.
    Our method produces the most natural results, and the weight maps (please refer to \ref{subsec:hybrid} for an explanation) reveal the parting lines correctly.
    Nearest neighbor (NN) interpolation shows unrealistic abrupt changes between patches.
    Bilinear (BL) interpolation has severe hair-head penetration issues near the parting line.
  }
  \label{fig:nhd-neural}
\end{figure*}

\begin{figure*}[t]
  \includegraphics[width=\linewidth]{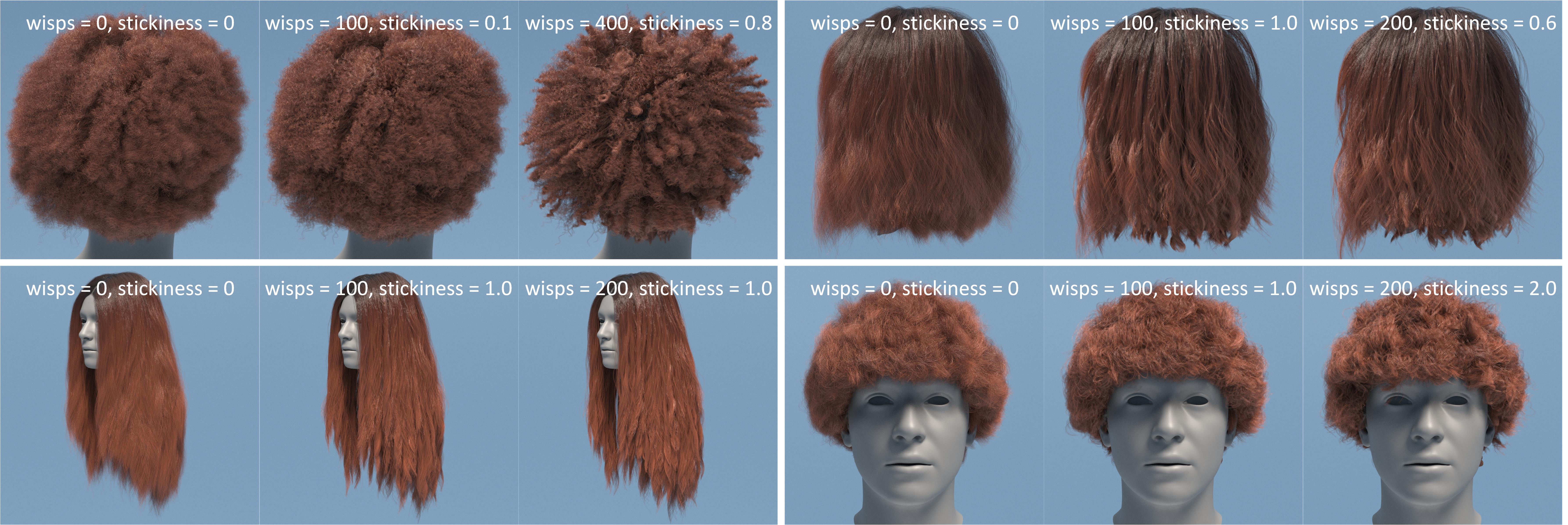}
  \caption{
    Our results after refinement.
    Each group is produced from the same output of the neural upsampler.
    The parameters effectively control the fine details without losing realism.
    Note the complex wisps structures, e.g. row 1 column 5, originates from the generated guide strands with high local variety.
  }
  \label{fig:nhd-refinement}
\end{figure*}

\begin{table}[t]
  \caption{
    Hair-head penetration rate of different interpolation methods.
    Our GAN-based neural upsampler avoids the aliasing artifacts in nearest neighbor interpolation but also keeps the parting line sharp, as indicated by a very low penetration rate.
  }
  \label{tab:nhd-pn}
  \begin{tabular}{cccccc}
    \hline
     & ours & NN & BL & full sup. & latent pred. \\
    \hline
    rate & 1.5\textperthousand\ & 0.0\textperthousand\ & 5.0\textperthousand\ & 2.1\textperthousand\ & 195.5\textperthousand\ \\
    \hline
  \end{tabular}
\end{table}

In the densification step, we utilize our neural upsampler to increase the number of strands by upsampling low-resolution latent-maps to higher resolution.
This is followed by a heuristic refinement process with user-defined parameters.
Please note that all the results presented in this step are based on the hairstyle-VAE's reconstruction and not ground truth guide strands.

We first evaluate the neural upsampler.
In \autoref{fig:nhd-neural}, we show a few representative hair models produced by the neural upsampler and compare them with alternative methods including nearest-neighbor (NN) and bilinear (BL) interpolation in both Euclidean and strand latent spaces.
To visualize the weight-maps intuitively, we factorize the weight for bilinear interpolation into individual guide strands and colorize each pixel based on the standard deviation (\textit{std}) of the weights.
Larger \textit{std} values indicate sharp transitions, while smaller \textit{std} values reflect smooth interpolations.
The emerging grid-like structure illustrates how interpolation is inhibited when close to the guides and parting lines, while being smooth otherwise.
To make artifacts more obvious, we remove all strands that penetrate the head mesh.
In the first row of \autoref{fig:nhd-neural}, our learned upsampler not only preserves the parting line but also avoids aliasing patterns observed in the NN interpolation (column 3 \& 5).
The presence of baldness in the bilinear interpolations (column 4 \& 6) indicates severe penetrations when interpolating strands from opposite sides of the parting line.
In the second row, our estimated parting line correctly ends before the fringe.
Additionally, we report the percentage of hairs that penetrate the head mesh after upsampling in \autoref{tab:nhd-pn}.
Our neural upsampler stands out with a very low penetration rate, indicating that it accurately identifies most hairstyle parting lines.
\revised{Please see \autoref{sec:more} for more results of the neural upsampler.}

In \autoref{fig:nhd-refinement} we demonstrate the final results after the refinement step.
Each group presents three hair models produced from the same output of the neural upsampler but with different user-defined parameters.
Note that the intricate wisp structures, such as the one shown in the first row, fifth column, are rarely seen in previous works.
The high fidelity of these structures is a result of our strand-level representation of hair models in the frequency latent space.
Please find more results in \autoref{sec:more}.

\subsection{Ablation Study}
\label{subsec:ablation}

In this section, we provide justifications for the important technical choices made in our approach.
We first explain why our frequency representation of strands leads to a better latent space compared to conventional spatial coordinates.
Then, we demonstrate the superiority of our GAN-based neural upsampler over other alternatives for strand-map upsampling.
Lastly, we highlight the suitability of our hierarchical structure for representing a hair model.

\paragraph{Strand VAE}
\label{subsec:ablation:svae}

\begin{figure}[t]
  \includegraphics[width=\linewidth]{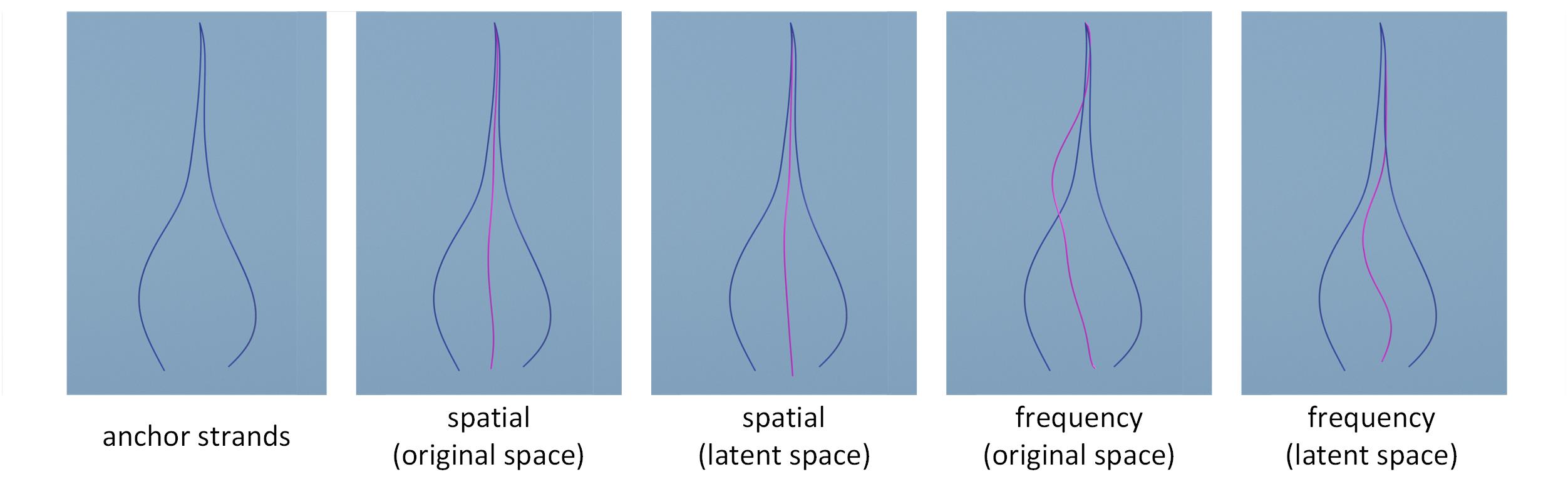}
  \caption{
    Averaging two curly strands (column 1, blue strands) based on spatial representations (column 2 and 3, magenta strand) results in straight strands in both original space and latent space.
    While frequency representation (column 4, magenta strand) preserves the curliness, averaging in the latent space (column 5, magenta strand) gives the most natural result.
  }
  \label{fig:spatial_vs_frequency_interp}
\end{figure}

\begin{figure}[t]
  \includegraphics[width=\linewidth]{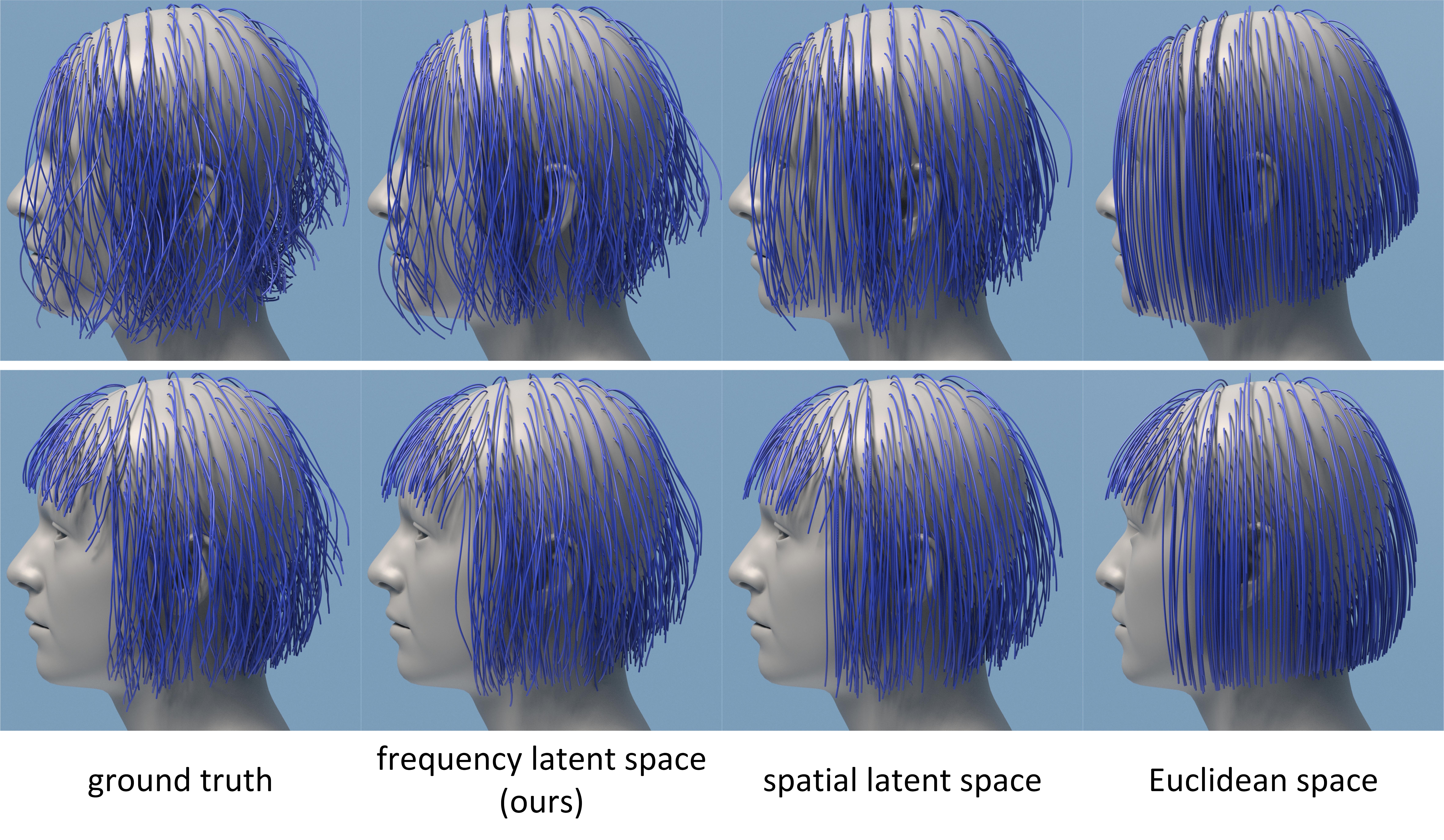}
  \caption{
    The frequency hairstyle-VAE (ours) achieves significantly better local variety while other alternatives with different strand representations or latent spaces suffer from over-smoothness.
  }
  \label{fig:spatial_vs_frequency_hae}
\end{figure}

\begin{table}[t]
  \caption{
    Positional error (pos. err.) and relative messiness (rel. mes.) for hairstyle-VAE variants.
    The unit is millimeter.
    Our method has better reconstruction error and also similar levels of hair messiness as the ground truth data.
    Other models are either over-smooth or less accurate.
  }
  \label{tab:hae-ablation}
  \begin{tabular}{cccccccc} 
    \hline
     & & \multicolumn{3}{c}{latent space} & & \multicolumn{2}{c}{original space} \\
     \cline{3-5} \cline{7-8}
     & & freq. (ours) & spat. & dense & & spat. & freq. \\
    \hline
    pos. err. & & 7.25 & 7.08 & 9.80 & & 7.07 & 8.47 \\
    \hline
    rel. mes. & & -0.05 & -0.11 & -0.23 & & -0.20 & -0.08 \\
    \hline
  \end{tabular}
\end{table}

Our method begins by constructing a compact latent space for strands, which effectively reduces the dimensionality while preserving high-fidelity shape information.
Among various shape features, we consider length and curliness to be the most crucial ones.
We opt to build the strand latent space in the frequency domain (\textit{frequency latent space}), motivated by the fact that the Fourier spectrum explicitly encodes both features.
In contrast, spatial coordinates do not directly represent curliness, so that the latent space thereon (\textit{spatial latent space}) is not always consistent with respect to curliness.
Consequently, two visually similar curly strands may be embedded far apart in this spatial latent space, while a straight strand and a curly strand may appear closer to each other.
Such counter-intuitive issues are avoided in the frequency latent space.
To validate this claim, we conduct additional experiments by training a separate strand-VAE model using spatial coordinates (\textit{spatial strand-VAE}) and subsequently a hairstyle-VAE based on it (\textit{spatial hairstyle-VAE}).
We refer to our main models in the frequency domain as \textit{frequency strand-VAE} and \textit{frequency hairstyle-VAE}.

We first present an intuitive explanation for our reasoning in \autoref{fig:spatial_vs_frequency_interp}.
We pick two curly hairs from our dataset and average them based on their spatial coordinates.
The resulting strand appears straight and loses its curliness (column 2).
Averaging in the spatial latent space leads to a similar outcome (column 3).
In contrast, if the averaging is performed on the frequency code $\mathcal{V}$, the curliness is better preserved (column 4).
The most meaningful result is obtained when the averaging is conducted in the frequency latent space (column 5).
This is because, in the frequency domain, the averaging is applied independently to the amplitudes and phases.
If both source strands have similar amplitudes but different phases, their mean will maintain the amplitude (i.e., curliness) while undergoing a phase shift.
Consequently, the frequency latent space aligns better with the human perception of hair shapes, where curly and straight strands are distinguished regardless of their spatial proximity.
\revised{Please find more results of strand interpolation in \autoref{sec:strand-interp}.}

This inherent structural distinction between the spatial and frequency strand latent spaces significantly influences the performance of the hairstyle-VAEs residing within them.
This impact is shown in \autoref{fig:spatial_vs_frequency_hae}, where the frequency hairstyle-VAE (2nd column) exhibits a high level of local variety that closely resembles the ground truth, while the spatial hairstyle-VAE (3rd column) produces overly smooth results.
In real hairstyles, it is quite common for adjacent strands to grow in opposite directions, resulting in interesting local variations.
However, the embedding within the spatial domain tends to diminish such local variety, whereas the frequency domain preserves it better.

To quantitatively assess local variation, we introduce the \textit{messiness} metric, defined as follows.
For each strand $i$, we calculate its mean deviation relative to its neighbors:
\begin{equation}
  \mathcal{D}_{i} = \frac{1}{|\mathcal{N}_i|}  \sum_{j \in \mathcal{N}_i} \frac{1}{N_s - 1} \sum_{k=1}^{N_s - 1} || \vd_{i, k} - \vd_{j, k} ||_2 ,
\end{equation}
where $\mathcal{N}_i$ represents the neighbors of strand $i$, and $d_{i, k}$ denotes the parent-relative displacement of vertex $k$ of strand $i$.
The messiness metric is defined as the mean $\mathcal{D}_i$ of all strands, characterizing the uniformity of a hair model.
A higher messiness value indicates greater local variety, while a lower value suggests a regular and smooth hairstyle.
We report the difference in messiness relative to the ground truth dataset ($0.428$mm), where a lower value indicates a smoother result.
As shown in \autoref{tab:hae-ablation} (rel. mes.), although our method produces a slightly more regular outcome compared to the ground truth, the spatial hairstyle-VAE exacerbates the gap, resulting in a doubling of the difference in messiness.

\revised{
  Notably, while positional encoding (PE) \cite{mildenhall2020nerf} shares certain high-level similarities with discrete Fourier transform (DFT), they are substantially different in our context.
  PE considers the coordinates individually and expands each scalar value to a high-dimensional vector, while DFT considers a sequence and transforms it into the Fourier domain with the same dimension.
  For comparison, we train another strand-VAE based on PE and a corresponding hairstyle-VAE, and observe similar local over-smoothness (relative messiness: -0.11) as the vanilla spatial representation.
  This is because PE does not considers the strand shape as a whole.
  The average pos. error (strand-VAE: 1.70mm, hairstyle-VAE: 6.89mm) is also similar.
}

\paragraph{Neural Upsampler}

\begin{figure}[t]
  \includegraphics[width=\linewidth]{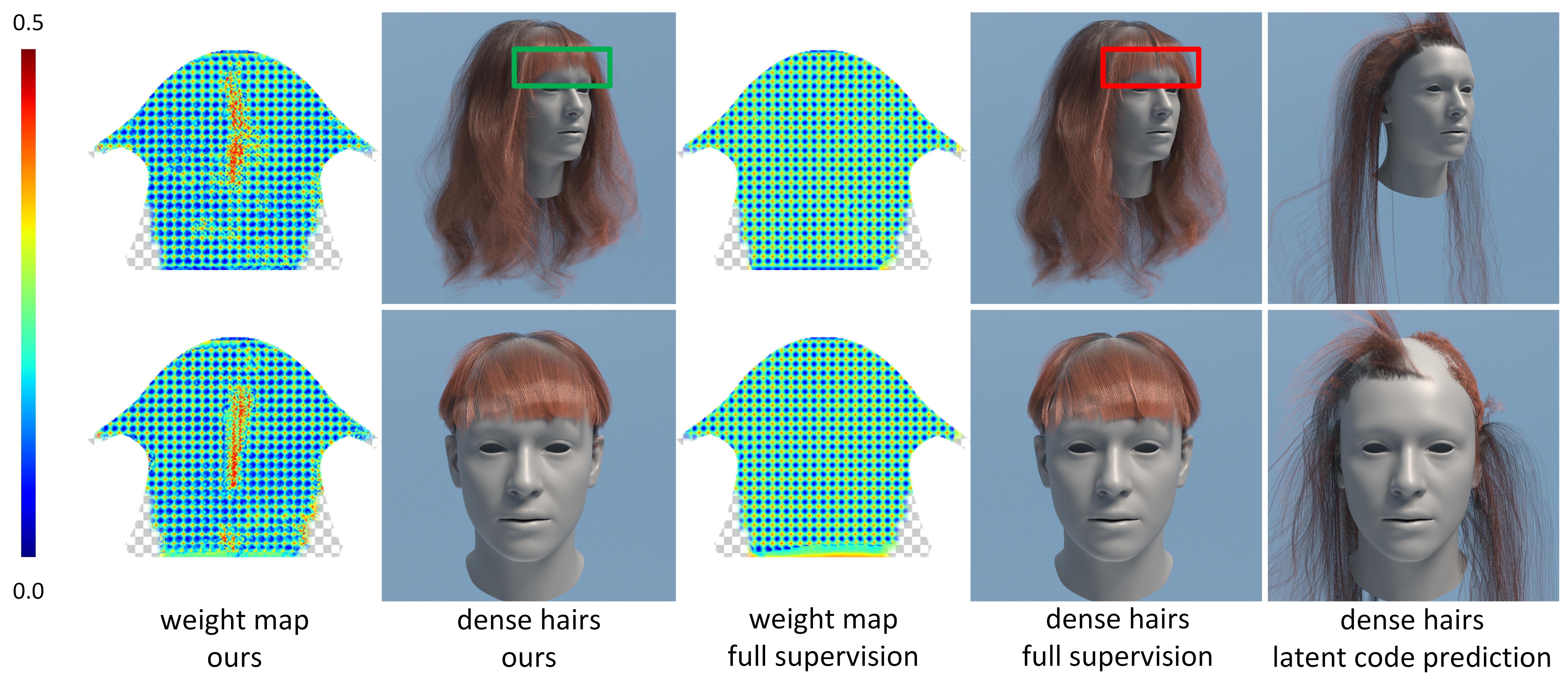}
  \caption{
    Densification results from different upsamplers.
    Our method (column 1 \& 2) works well on diverse hairstyles.
    Training with full supervision (column 3 \& 4) is comparable to bilinear interpolation.
    Directly learning strand geometry fails to converge (column 5).
  }
  \label{fig:nhd_ablation}
\end{figure}

The challenge of hair densification is two-fold.
Firstly, the sparse guide strands only provide a general depiction of the overall hairstyle, making the mapping to dense hair strands indeterminate without unique ground truth.
Secondly, humans perceive hairstyles holistically rather than focusing on individual strands.
Per-strand supervision becomes impractical when dealing with as many as $25$K strands.
Therefore, we choose to adopt a GAN model for perceptual supervision.

To highlight the advantages of our GAN-based neural upsampler, we train an alternative model using full supervision.
The ground truth data is generated by downsampling the hair models in our database.
However, we find that this model cannot be effectively optimized and only converges to a local optimum that closely resembles bilinear interpolation. This occurs because the model attempts to learn a deterministic mapping that does not exist.
The weight maps and qualitative results of this fully supervised model exhibit similar unnatural artifacts as bilinear interpolation and lack awareness of parting lines, as shown in \autoref{fig:nhd_ablation} (column 3 \& 4).

We choose to predict interpolation weights instead of strand geometry due to the greater constraints imposed on the weights.
This leads to more stable training for the fragile GAN model.
As an ablation analysis, we train another model that directly predicts strand latent codes using a discriminator loss.
This model fails to converge to a valid solution and only generates meaningless strands, as shown in \autoref{fig:nhd_ablation} (column 5) and \autoref{tab:nhd-pn} (column \textit{latent pred.}).

\paragraph{Hierarchical Structure}
Representing hairstyles with a hierarchical structure is one of our main design choices.
We demonstrate the necessity of each level in the following analysis.
At the strand level, encoding strands into a low-dimensional latent space effectively simplifies the generation task.
To verify this, we conduct an ablation study by removing the strand-level abstraction. We train two alternative hairstyle-VAE models that directly take strand geometries from the original space.
These models are denoted as \textit{ori. spat.} (strands represented as spatial gradients $\bar{\mathcal{S}}$ in the original space) and \textit{ori. freq.} (strands represented as frequency codes $\mathcal{V}$ in the original space).
They share the same network structure as our standard hairstyle-VAE, except for different input and output dimensions.
The quantitative evaluations are reported in \autoref{tab:hae-ablation}.
Although the ori. spat. variant achieves a slightly lower positional error, it suffers from over-smoothness, as shown in \autoref{fig:spatial_vs_frequency_hae} (4th column).
On the other hand, the ori. freq. variant produces a considerably worse reconstruction error, reported in \autoref{tab:hae-ablation}.

At the hairstyle level, we utilize a set of sparse guide strands to describe a hair model instead of dense strands.
Using guide strands as an intermediate descriptor is crucial for reducing the complexity of the task.
In our case, dense strands are represented by a latent-map of shape $216 \times 288 \times 65$, containing more data than a high-resolution ($1280 \times 1024$) RGB image.
Compressing such a large amount of data into a low-dimensional vector poses a challenging task in its own right, let alone that no control is provided to the user in such an end-to-end model.
For the ablation study, we remove the sparse guide strands level modeling and train an alternative hairstyle-VAE model that learns the mapping directly from dense strands to a single latent hairstyle vector, denoted as \textit{dense} in \autoref{tab:hae-ablation}.
Despite having $1.6$ times more parameters than the combined hairstyle-VAE and neural upsampler, this model still has worse reconstruction error and struggles with local over-smoothness.

\subsection{Application: Quasi-Static Simulation}
\label{subsec:simulator}

As an example application based on the latent representation of a hairstyle, we introduce a quasi-static simulator.
Regarding the shape of a hairstyle with the straight head pose as the rest state, the neural simulator estimates hair deformation driven by different head poses.
We model hair in the head coordinate frame, where the head is fixed while the direction of gravity can vary.
Thus, the task is formulated as predicting strand deformation given a specific gravity direction.

Our quasi-static simulator is a neural network that takes as input a reference latent code $\vl_r\in\mathbb{R}^{D_s}$ of the target strand under standard gravity ($-y$ direction), the position of the strand root $\vp_1 \in \mathbb{R}^3$, and the new direction of gravity $\vh\in\mathbb{R}^3$.
The output is the latent code of the deformed strand.
We design the network to be of $33$ fully-connected layers with $1024$ hidden units and residual connections, optimized using the L2 loss of the target latent codes.
The training data is built from \textsc{GroomHair} where the head is randomly tilted and the equilibrium state of the guide strands is simulated using \textit{Houdini}.
The deformed strand geometries are embedded into the strand latent space with the pre-trained strand-VAE.

\begin{figure}[t]
  \includegraphics[width=\linewidth]{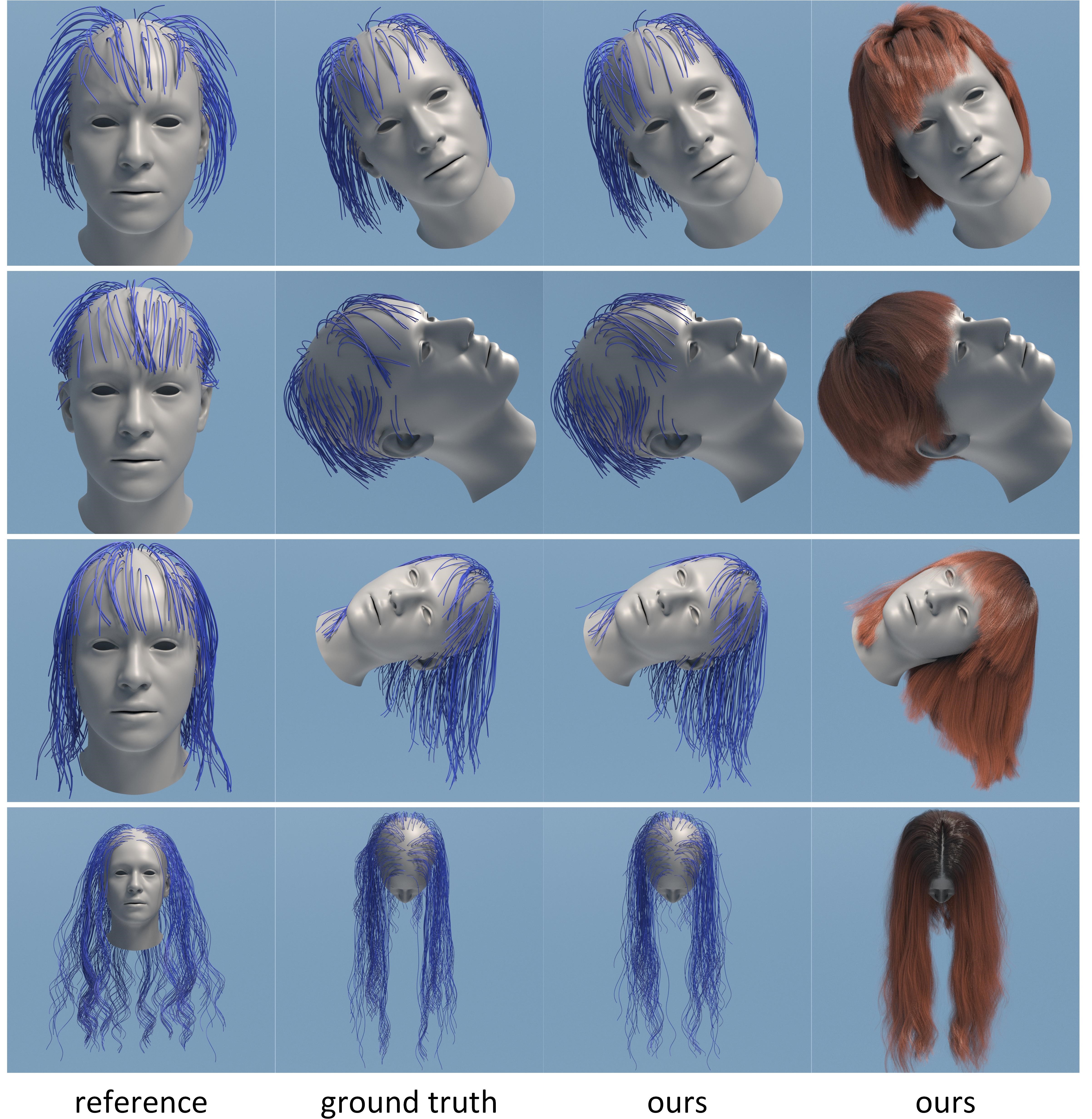}
  \caption{
    Our neural simulator gives plausible estimations for diverse hairstyles and head poses.
    It runs in real-time for up to 3K hairs, while conventional industrial simulators may take minutes.
  }
  \label{fig:nss-test}
\end{figure}

As shown in \autoref{tab:eval-quant} the average error of the neural simulator is only $9$mm.
In \autoref{fig:nss-test}, we provide qualitative results where the predicted deformation closely matches the ground truth.
Compared to conventional methods, the neural simulator is faster by 3 magnitudes (\autoref{tab:runtime}), but still achieves satisfying quality, suggesting that the latent representation can support complex downstream tasks.

\section{Conclusion}
\label{sec:conclusion}

In this paper, we present the first generative hair model capable of automatically synthesizing diverse hairstyles.
We demonstrate the effectiveness of embedding hairstyles into latent spaces with significantly fewer parameters through hierarchical decomposition.
The strand latent space, based on frequency components, reduces dimensionality while preserving fidelity.
The hairstyle latent space is well-constrained for generating guide strands.
The neural upsampler effectively densifies guide strands into dense hairs, and the heuristic refinement process produces realistic final results with user control.

\begin{figure}[t]
  \includegraphics[width=\linewidth]{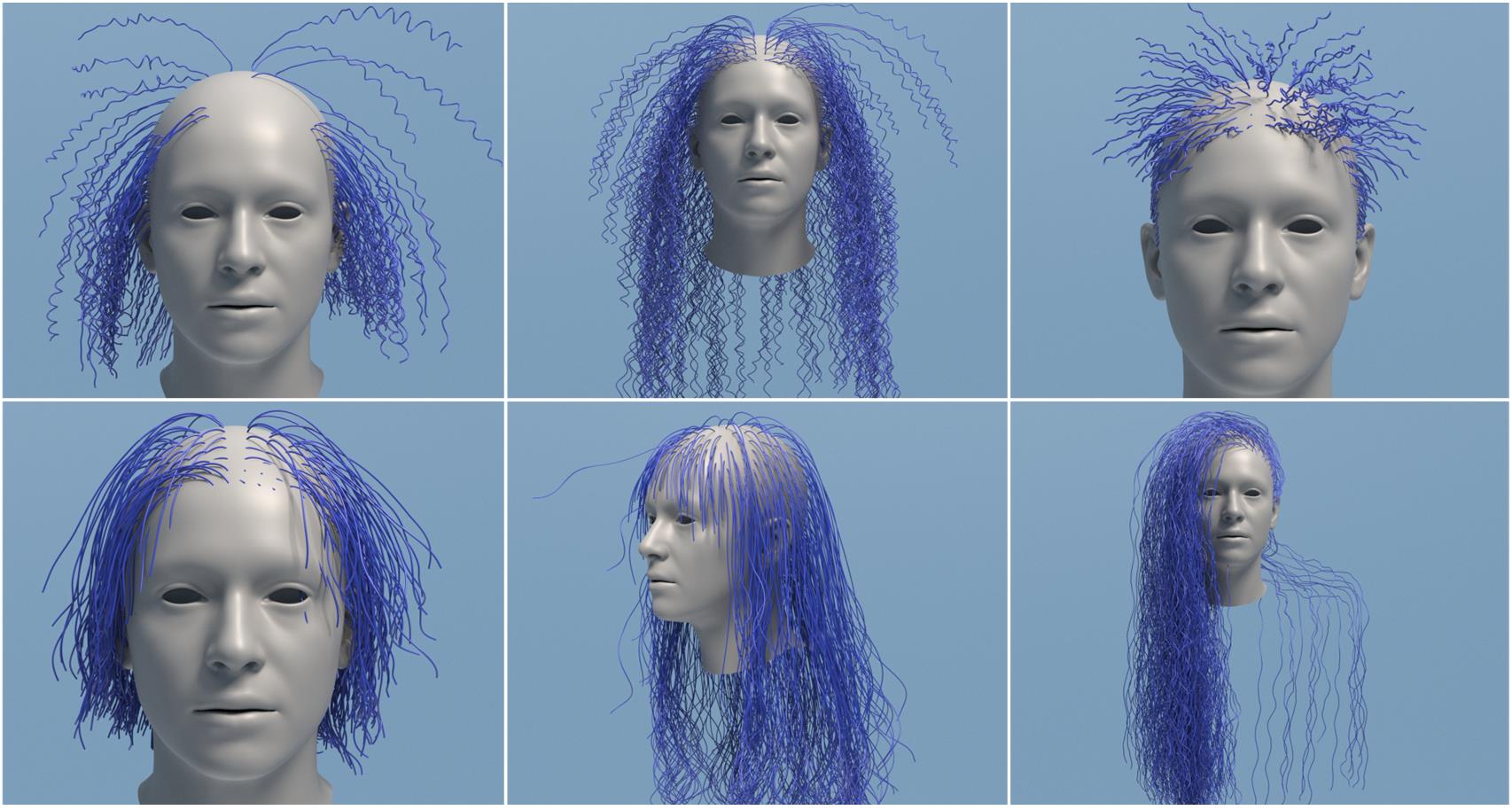}
  \caption{
    Failure cases that are generated from random sampling.
    Top row: unnatural hairstyles do not conform to human aesthetics.
    Bottom row: physical artifacts such as penetration and flying long strands.
  }
  \label{fig:failure}
\end{figure}

\paragraph{Limitations.}
First, the generation capability of our model is inherently bounded by the diversity of the dataset.
The majority of the dataset comprises everyday hairstyles, while certain styles like braids are not adequately represented.
Second, the entire system is trained on a specific head shape, embedded within the network weights.
While the UV parameterization allows for adaptation of generated hair models to different head shapes to a certain extent, penetrations may still occur since no explicit information about the head mesh is provided to the system.
Lastly, our current system does not explicitly consider physical attributes.
The empirically devised wisp formation and penetration refinement steps lack a solid physical foundation, and the neural simulator infers hair deformation solely based on the rest shape and pre-defined gravity.
\revised{In \autoref{fig:failure} we show a few failure cases from random generation.}

\paragraph{Applications}
The primary application of this work is automatic hairstyle generation.
With the well-structured hairstyle latent space and the refinement steps, our method is capable of generating hair models that go beyond the training data to a certain extent.
The hierarchical modularization of the pipeline also allows for human involvement, where artists can edit the generated guide strands before the automatic densification step or fine-tune the heuristic parameters for more precise control during the refinement stage.
Additionally, we anticipate that the strand and hairstyle latent space can serve as reliable priors for hair geometry acquisition, leading to a hair capturing system that starts by optimizing the hairstyle and strand latent codes to fit the observed data.

\begin{acks}
  We thank the anonymous reviewers for their insightful comments and suggestions, Denis Zen for modeling the hairstyles in the dataset, Shunsuke Saito for helping with the comparison, Xinyu Yi for proofreading, Daoye Wang for code review, and Erroll Wood and Chenglei Wu for fruitful discussions.
\end{acks}

\bibliographystyle{ACM-Reference-Format}
\bibliography{reference}

\newpage
\appendix
\section{Additional Results}
\label{sec:more}

\begin{figure}[b]
  \includegraphics[width=\linewidth]{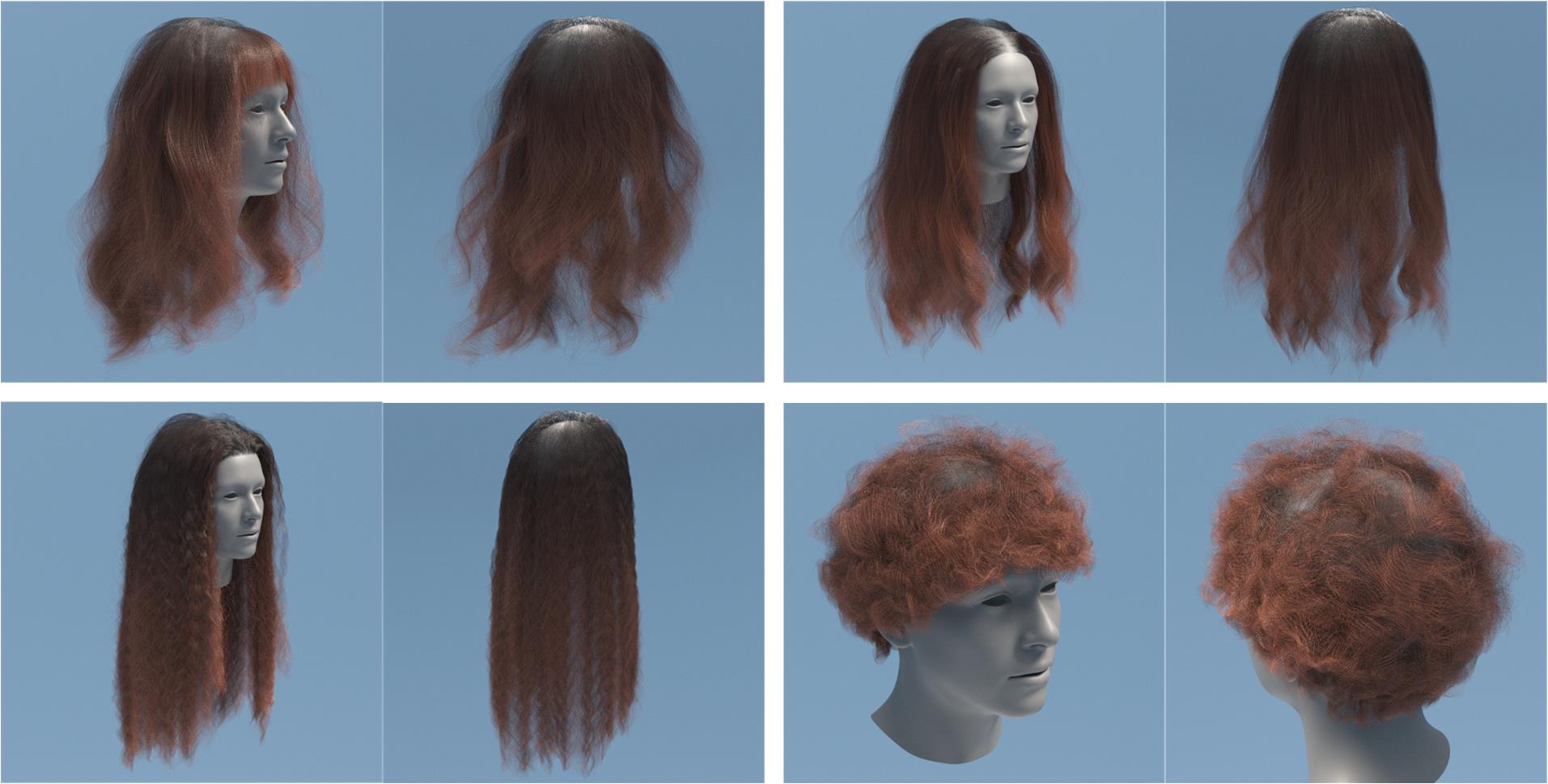}
  \caption{
    \revised{Raw outputs from our neural upsampler.}
    \revised{It effectively populates the guide hairs, preserves the shape, keeps the parting lines, and avoids unnatural patterns.}
  }
  \label{fig:nhd-raw-output}
\end{figure}

\begin{figure*}
  \includegraphics[width=\linewidth]{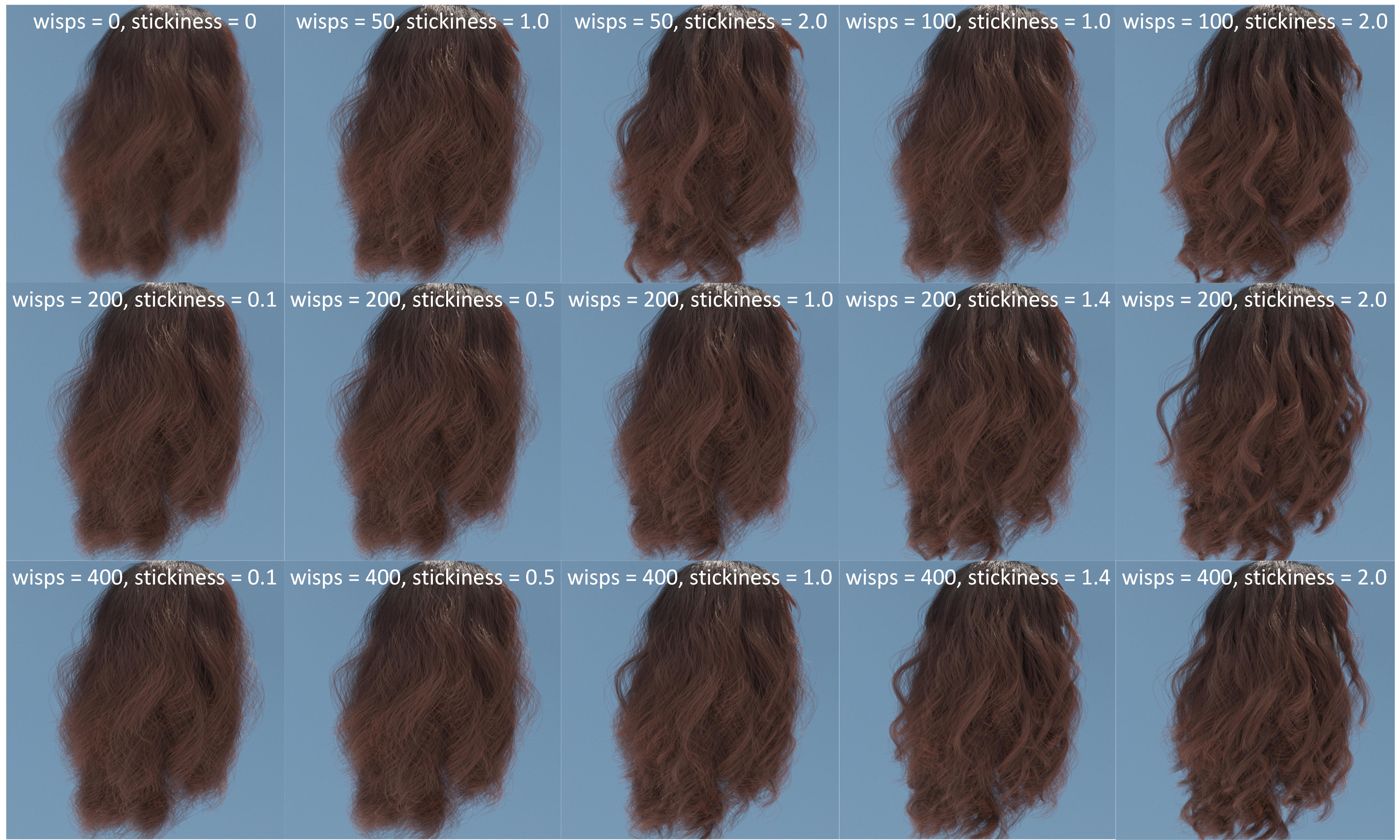}
  \caption{
    Hair models from the same guide strands but varied user parameters.
  }
  \label{fig:nhd-params}
\end{figure*}

\revised{In \autoref{fig:nhd-raw-output}, we present the raw outputs of the neural upsampler to evaluate its densification power as an individual module.}
In \autoref{fig:nhd-params} we demonstrate various combinations of the user parameters.

\begin{figure}[h]
  \includegraphics[width=\linewidth]{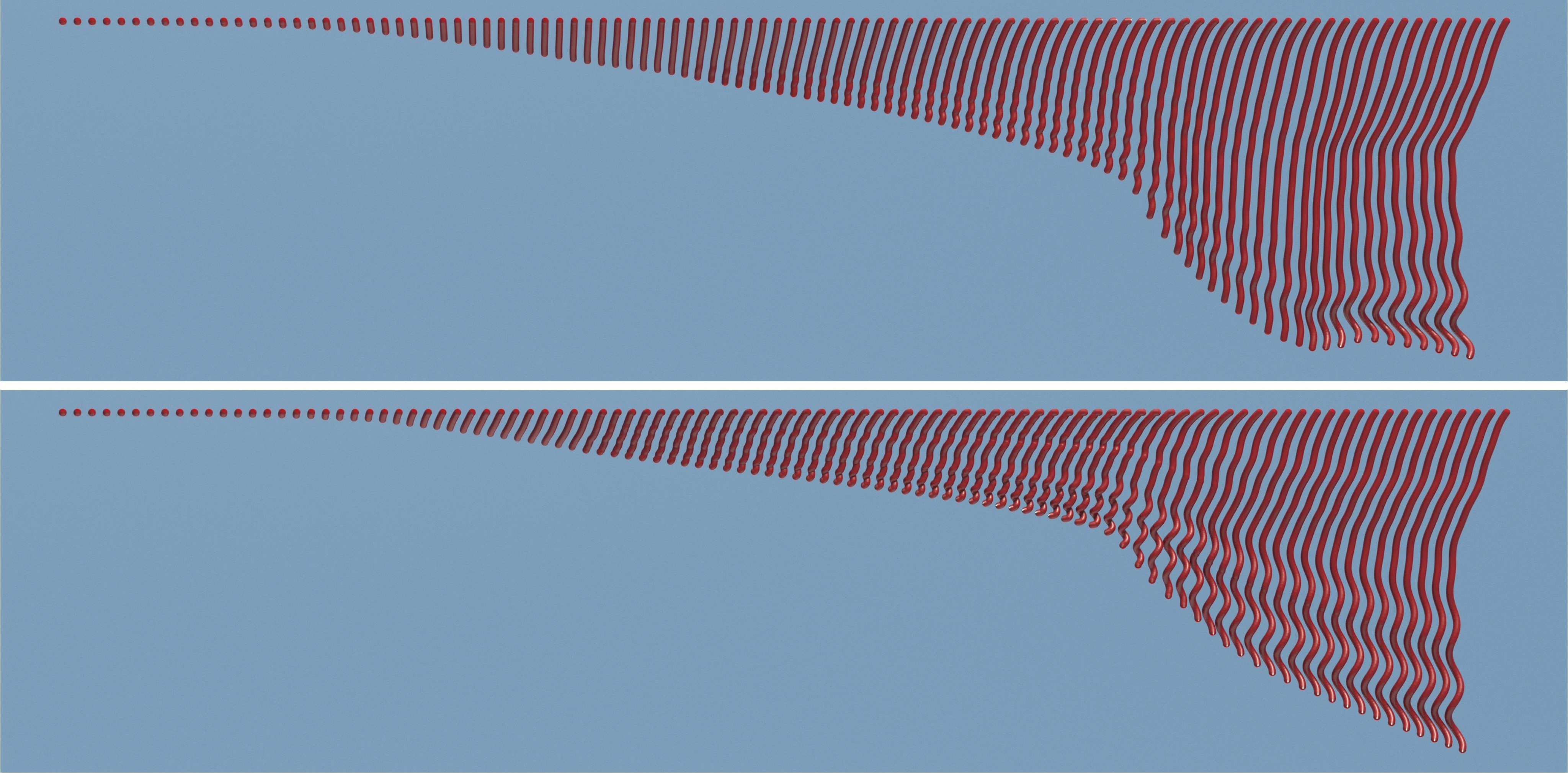}
  \caption{
    \revised{
      Interpolation trajectories between a short strand and a long wavy strand in different latent spaces.
      Top: spatial latent space; Bottom: frequency latent space.
      The frequency representation preserves the curvature better.
    }
  }
  \label{fig:strand-interp-seq}
\end{figure}

\section{Strand Interpolation}
\label{sec:strand-interp}

In \autoref{fig:strand-interp-seq}, we compare interpolation trajectories of two distinct strands in different latent spaces.
Interpolation in the spatial latent space demonstrates unstable curvature changes, while interpolation in the frequency latent space aligns better with human perception.

Moreover, we quantitatively examine the jittering effect during strand interpolation with different representations.
We randomly select $10$K strand paris from the test set and perform interpolation between them with $98$ intermediate sample points.
For each interpolation sequence, we then compute the \textit{jitter} metric~\cite{flash1985coordination}, defined as the third derivative of position by assuming the time interval is $1$ second, where a smaller value means a smoother transition.
As shown in \autoref{tab:jitter}, the transition in the frequency latent representation is as smooth as the spatial-latent representation with better curvature preservation.
This suggests that the frequency-latent space is better structured.

\section{Penetration Refinement}
\label{sec:penetration}

Our system is based on the same head geometry and the detailed head shape is not explicitly considered.
Although the shape information is partially baked into the weights of the neural networks, penetrations still happen occasionally.
To mitigate the penetration artifacts and preserve the hair structure as much as possible, the following refinement is performed.
For each strand in the Euclidean space, we traverse its vertices from root to tail.
At vertex $\vp_i$, we check if the vertex ahead, $\vp_{i+k}$, is within the head mesh using a pre-computed signed distance field.
If $\vp_{i+k}$ is inside the mesh, we compute the minimal rotation angle $\theta$ that pushes $\vp_{i+k}$ out of the mesh along the normal $\vn$ of the nearest surface with $\vp_{i}$ as the rotation pivot and $\vb = (\vp_{i+k} - \vp_{i}) \times \vm$ as the axis.
The vertices $\vp_{j} (j >= i)$ are rotated by $\theta_i = \theta * \delta^{j - i - k / 2}$.
After traversing all the vertices, we remove the strands that still penetrate the head mesh.
We empirically set $k = 20$, $\delta = 0.9$ for all experiments.

\begin{table}[t]
  \caption{
    \revised{The \textit{jitter} metrics of strand interpolations in different domains.}
    \revised{While vanilla Euclidean interpolation is most smooth, our latent representation in the frequency space is similarly smooth as the spatial one.}
  }
  \label{tab:jitter}
  \begin{tabular}{cccc}
  & Euclidean & spatial-latent & frequency-latent \\
  \hline
  jitter ($\mathrm{mm/s^3}$) & 0.69 & 0.89 & 0.97 \\
  \hline
  \end{tabular}
\end{table}

\begin{table}[t]
  \caption{
    Detail structure of the hairstyle-VAE model (encoder part).
    The decoder is symmetric.
    The residual connections are between layers 1 \& 3, 4 \& 6, and 7 \& 9 using bilinear downsampling.
    Layer 11 gives the final output with 1024 channels, half of which represents the latent vector while the other half is the log variation used for the reparameterization trick in VAE training.
    Layer 12 is used in the residual connection between layer 1 and 3 to align the number of channels after downsampling.
  }
  \label{tab:hae-network}
  \begin{tabular}{ccc}
  layer number & input size         & convolution                \\
  \hline
  1            & $24 \times 32 \times 65$   & (1, 1, 65, 2048, 1)    \\
  \hline
  2            & $24 \times 32 \times 2048$  & (3, 3, 2048, 2048, 2)  \\
  \hline
  3            & $12 \times 16 \times 2048$ & (1, 1, 2048, 512, 1)   \\
  \hline
  4            & $12 \times 16 \times 512$  & (1, 1, 512, 2048, 1)    \\
  \hline
  5            & $12 \times 16 \times 2048$  & (3, 3, 2048, 2048, 2)  \\
  \hline
  6            & $6 \times 8 \times 2048$ & (1, 1, 2048, 512, 1)   \\
  \hline
  7            & $6 \times 8 \times 512$  & (1, 1, 512, 2048, 1)    \\
  \hline
  8            & $6 \times 8 \times 2048$  & (3, 3, 2048, 2048, 2)    \\
  \hline
  9            & $3 \times 4 \times 2048$  & (1, 1, 2048, 512, 1)    \\
  \hline
  10           & $3 \times 4 \times 512$  & (3, 4, 512, 1024, 1)    \\
  \hline
  11           & $1 \times 1 \times 1024$  & (1, 1, 1024, 1024, 1)    \\
  \hline
  \hline
  \textit{12}  & $12 \times 16 \times 65$  & (1, 1, 65, 512, 1)    \\
  \hline
  \end{tabular}
\end{table}

\begin{table}[H]
  \caption{
    Detail structure of the neural upsampler (generator part).
    The discriminator is the same except the first and last layers.
    The residual connections are between layers 4 \& 6, 7 \& 9, and 10 \& 12.
  }
  \label{tab:nhd-network}
  \begin{tabular}{ccc}
  layer number   & input size         & convolution                \\
  \hline
  1            & $216 \times 288 \times 364$   & (1, 1, 128, 364, 1)    \\
  \hline
  2            & $216 \times 288 \times 128$   & (13, 13, 128, 128, 1)    \\
  \hline
  3            & $216 \times 288 \times 128$   & (1, 1, 128, 128, 1)    \\
  \hline
  4            & $216 \times 288 \times 128$   & (1, 1, 128, 128, 1)    \\
  \hline
  5 - 6        & \multicolumn{2}{c}{same as layer 2 - 3}                 \\
  \hline
  7 - 9        & \multicolumn{2}{c}{same as layer 4 - 6}                 \\
  \hline
  10 - 11      & \multicolumn{2}{c}{same as layer 4 - 5}                 \\
  \hline
  12           & $216 \times 288 \times 128$   & (1, 1, 128, 5, 1)    \\
  \hline
  \end{tabular}
\end{table}

\section{Network Structures}
\label{sec:network-detail}

In \autoref{tab:hae-network} and \autoref{tab:nhd-network} we provide the detailed structure of our hairstyle-VAE and neural upsampler, respectively.
The input size is formatted as $h \times w \times c$ where $h$, $w$, and $c$ are height, width, and channels.
The convolution is formatted as (kernel height, kernel width, input channels, output channels, stride).

\section{Qualitative Comparison with VHV}
\label{sec:vhv-comp}

\begin{figure}[h]
  \includegraphics[width=\linewidth]{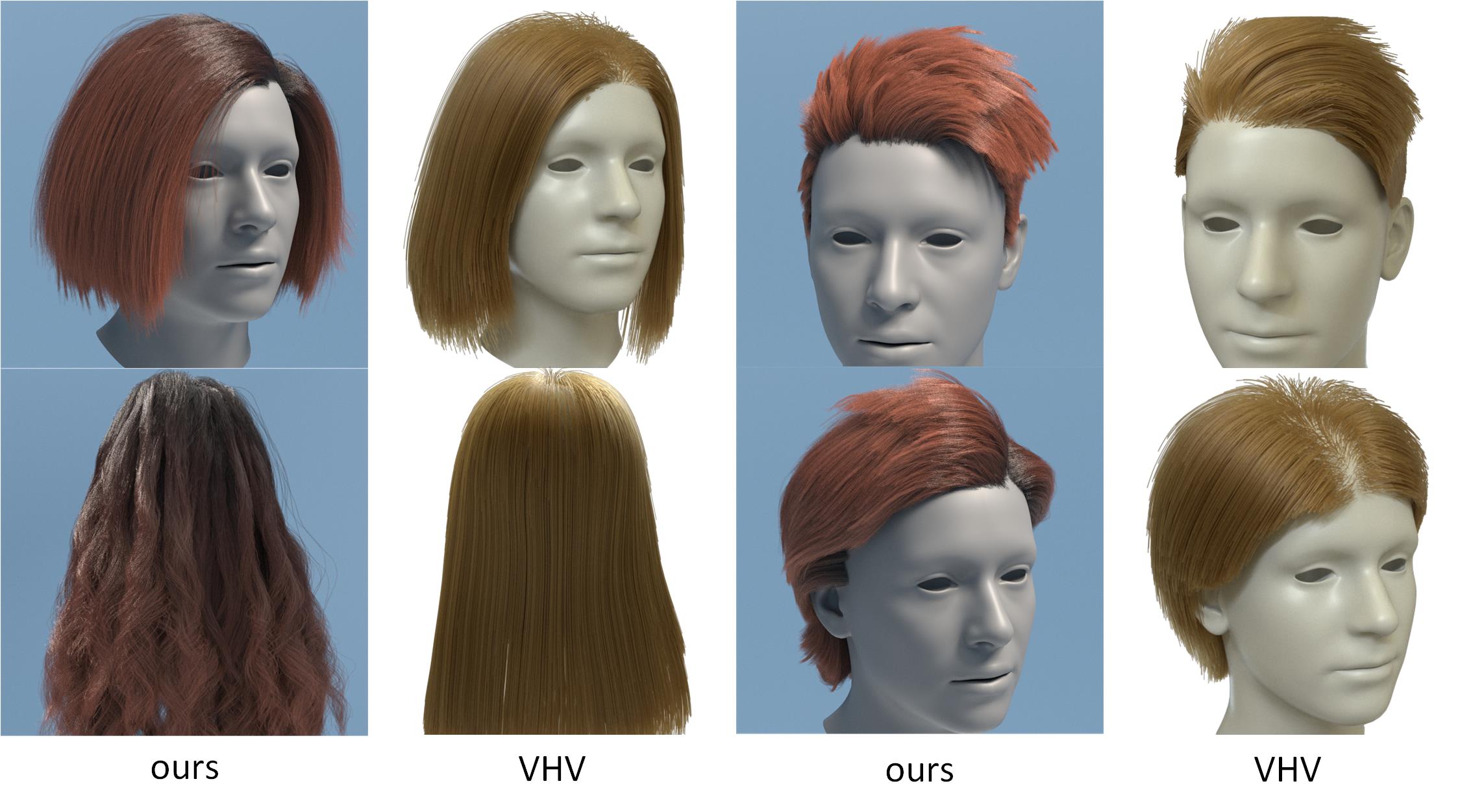}
  \caption{
    \revised{
      VHV reconstruction results of the rendered hair models generated by our method.
      While the overall hairstyles are successfully recovered, the reconstructions suffer from over-smoothing and lack of high-fidelity details.
    }
  }
  \label{fig:vhv-comp}
\end{figure}

\revised{
  To qualitatively compare our model with the seminal work of VHV \cite{saito20183d}, in \autoref{fig:vhv-comp}, we present VHV reconstruction results of the rendered hair models generated by our method.
  While VHV can recover the overall hairstyle well, some strand- and wisp-level details are missing, partially due to the limited capability of volumetric representations.
  Furthermore, some styles, such as the bottom-right hair model, are beyond the coverage of VHV.
  Note that this is not a strictly fair comparison as VHV is reconstructing the hair model only from a single-view image, and the face alignment step in the original method does not work on our non-photorealistic face rendering.
}

\section{Dataset}

\label{sec:dataset-list}
\begin{figure*}[t]
  \includegraphics[width=\linewidth]{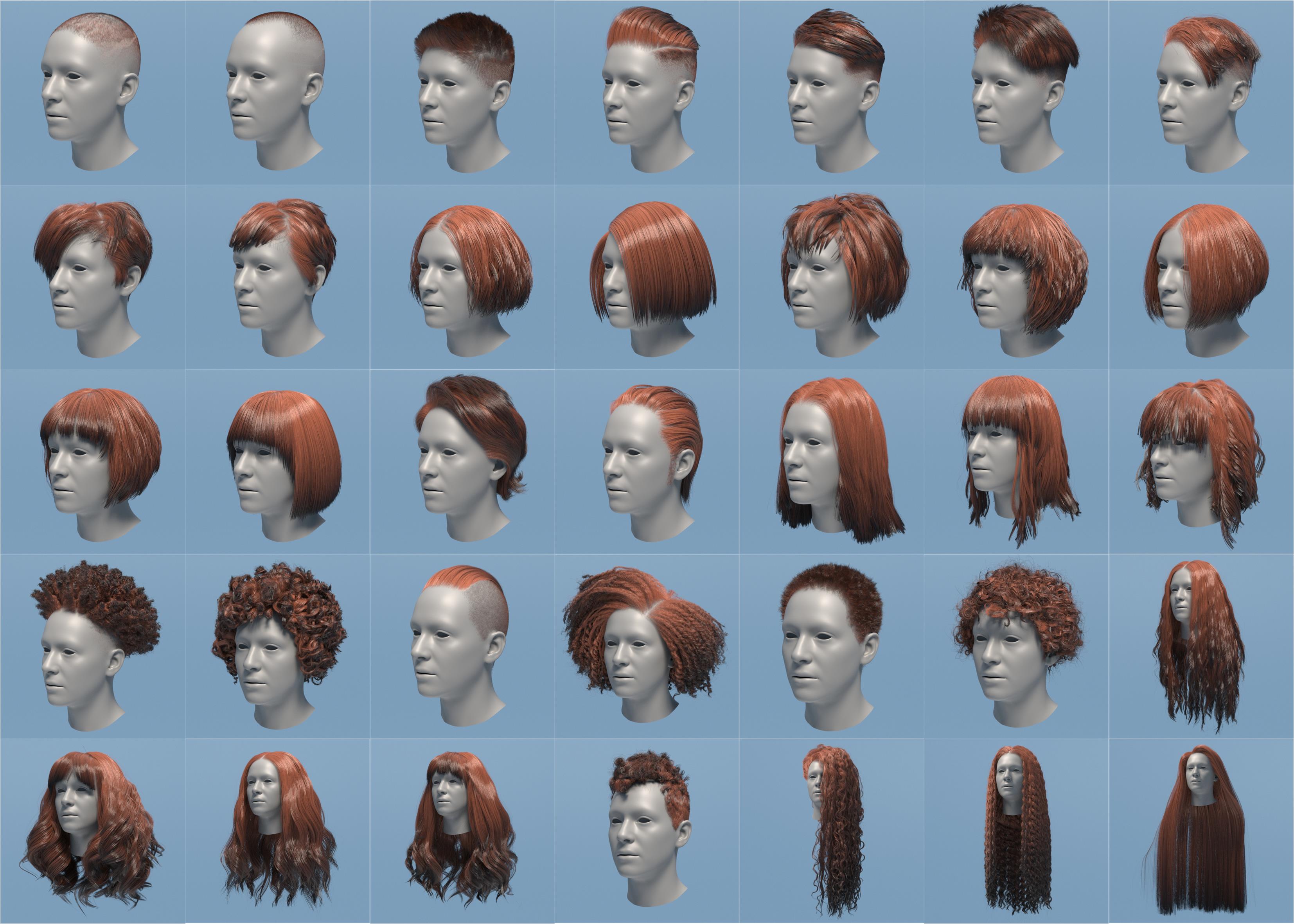}
  \caption{
    \revised{All 35 base hairstyle categories of our dataset.}
  }
  \label{fig:hairstyle-list}
\end{figure*}

\revised{
In \autoref{fig:hairstyle-list} we visualize the complete list of $35$ base hairstyle categories that constitute our dataset, which covers a wide variety of hairstyles.
}

\end{document}